\documentclass[10pt, a4paper]{article}
\usepackage{graphicx}%
\usepackage{multirow}%
\usepackage{amsmath,amssymb,amsfonts}%
\usepackage{amsthm}%
\usepackage{mathrsfs}%
\usepackage[title]{appendix}%
\usepackage{xcolor}%
\usepackage{textcomp}%
\usepackage{manyfoot}%
\usepackage{booktabs}%
\usepackage{algorithm}%
\usepackage{algorithmicx}%
\usepackage{algpseudocode}%
\usepackage{listings}%
\usepackage{enumitem}
\usepackage{epstopdf}
\usepackage[utf8]{inputenc}
\usepackage{authblk}
\usepackage{latexsym}
\usepackage{mathtools}
\usepackage{rotating}
\usepackage{tabularx}
\usepackage{natbib}
\usepackage{lineno} % for line numbering
\usepackage{aas_macros}
 \usepackage[
left=1.5cm,
right=1.5cm,
top=2.00cm,
bottom=2.00cm
]{geometry}
\usepackage{titlesec}
\usepackage{abstract}
\usepackage{bm}
\usepackage{svg}
\usepackage{array}
\DeclareUnicodeCharacter{02C8}{'}
\usepackage[utf8]{inputenc}
\usepackage{hyperref}

\title{\Huge\textbf{The ESA Meerkat Asteroid Guard: a monitoring service for imminent impactors}}

\author[1]{Charlie Drury \thanks{Email: \texttt{cjpdrury@gmail.com}}}
\author[1,2]{Francesco Gianotto}
\author[1,3]{Marco Fenucci} 
\author[1,2]{Laura Faggioli}
\author[4]{Michael Fr{\"u}hauf}
\author[1,5]{Juan Luis Cano}
\author[1,2]{Marco Micheli}
\author[6,7]{Francisco Oca\~{n}a}
\author[1,3]{Dario Oliviero}
\author[1]{Luca Conversi}
\author[1,5]{Richard Moissl} 
\author[4]{Detlef Koschny}

\affil[1]{Planetary Defence Office / NEO Coordination Centre, ESA ESRIN, Largo Galileo Galilei 1, Frascati, 00044, RM, Italy}

\affil[2]{Starion Italia, Via di Grotte Portella 28, Frascati, 00044, RM, Italy}

\affil[3]{Deimos Italia s.r.l.,  Via Alcide De Gasperi, 24, San Pietro Mosezzo, 28060, NO, Italy}

\affil[4]{Lunar and Planetary Exploration, Technical University of Munich,  Lise-Meitner-Str. 9, Ottobrunn, 85521, Germany}

\affil[5]{Planetary Defence Office, ESA ESOC,   Robert-Bosch-Stra{\ss}e 5, Darmstadt, 64293, Germany}

\affil[6]{Planetary Defence Office, ESA ESAC,  Villafranca del Castillo, Bajo del Castillo, Madrid, 28692, Spain}

\affil[7]{Deimos Spain S.L.U.,  Ronda de Poniente, 19, Tres Cantos Madrid, 28760, Spain}

\date{\today}

\begin{document}

\maketitle

\begin{abstract}
We present the Meerkat Asteroid Guard, an imminent impactor warning service developed and maintained by the European Space Agency’s Near-Earth Object Coordination Centre (NEOCC). The software uses the method of systematic ranging to perform orbit determination on tracklets in the Near-Earth Object Confirmation Page (NEOCP), which typically have short observational arcs. Fitted orbits are propagated to determine the likelihood of an impact with Earth. In addition, magnitude fitting and Monte Carlo sampling are performed to estimate the object’s size, possible impact locations and times, and suggest a best telescope pointing for object follow-up. A set of object scores are produced from computed posterior probabilities across the grid, giving a statistical description of the object’s orbital and physical characteristics. The scores are packaged with several informative plots in an email alert, which is sent to Meerkat subscribers in the event of a significant impact probability, close approach, or other scientifically interesting event. The highlights of the five years of Meerkat’s operational service are presented, including the successful warnings for all of the past six imminent impactors discovered before impact and several interesting close approaches.

\end{abstract}

\section{Introduction}\label{s:intro}
An \emph{imminent impactor} is a near-Earth object (NEO) that is discovered a few days before impacting Earth. Imminent impactors are expected to be smaller objects, with diameters smaller than 50 m. This is because they may only be detected when close to Earth and are much more numerous than larger objects \cite{granvik-etal_2018, nesvorny-etal_2024}. While the damage caused by the impact of such small asteroids is not at a global level, they can still produce local damage to people and infrastructure, especially if they fall over a populated area. Indeed, it is estimated that decameter-size asteroids are the most likely to cause significant ground damage in the near future \citep{chow-etal_2025, boslough-etal_2015, gi-etal_2018}. Such an event occurred recently in February 2013, when a small asteroid with a diameter of about 20 m fell close to the Russian city of Chelyabinsk \citep{popova2013}. The atmospheric entry and subsequent fragmentation of the parent body generated a shock wave which shattered glass in the surrounding area, causing injury to more than 1600 people \cite{kartashova-etal_2018}. The total energy released by the impact was estimated to be 500 $\pm$ 100 kT TNT equivalent \citep{borovika2013}.

The asteroid which fell over Chelyabinsk was not spotted before impact, motivating the planetary defence community to advance the efforts in the searching and monitoring of imminent impactors. By 2013 both the Catalina Sky Survey \citep[CSS,][]{fuls-etal_2023} and the Panoramic Survey Telescope and Rapid Response System \citep[Pan-STARRS,][]{denneau-etal_2013} were active, and fully dedicated to the search of larger asteroids. In addition to these, NASA also funded the Asteroid Terrestrial-impact Last Alert System \citep[ATLAS,][]{tonry-etal_2018}. ATLAS is a network of four telescopes (two located in Hawaii, one in South Africa, and one in Chile) with a field of view of 7.4 deg$^2$, able to scan the whole sky in a single night down to magnitude 20. In addition, a second-generation ATLAS telescope has been recently installed in Tenerife \citep{licandro-etal_2023}. The system is therefore specifically suitable for the discovery of imminent impactors.
To complement the American surveys, the European Space Agency (ESA) conceived and funded the Flyeye telescope \citep{conversi-etal_2021, diprima2023, arcidiacono2024}. This instrument has an innovative design which results in a very large field of view of $\sim$44 deg$^2$ which, coupled with a highly automated equatorial mount, permits the entire observable sky to be scanned every few nights down to approximately magnitude 20. This survey is also specifically designed to search for imminent impactors. Currently, the first Flyeye telescope is undergoing a testing phase at the Space Geodesy Center of the Italian Space Agency (ASI) in Matera (Italy), and is planned to be moved to Mount Mufara in Sicily in 2026, where it will start nominal operations. Another Flyeye telescope is planned to be developed, and installed in the southern hemisphere \citep{fohring2024}.  

Not only are imminent impactors difficult to discover, further challenges emerge from the point of view of orbit determination and impact prediction. Such objects need to be monitored right at the discovery time, when the observational arc is typically very short. Classical orbit determination methods \citep{milani-gronchi_2009} are not suitable for short-arc orbit determination, motivating other techniques to be developed \citep{farnocchia-etal_2015c}. Today, there are four operational monitoring systems for imminent impactors: the Meerkat Asteroid Guard by ESA; the Scout system \citep{farnocchia-etal_2015c} by NASA JPL\footnote{\url{https://cneos.jpl.nasa.gov/scout/}}; the NEOScan \citep{spoto-etal_2018, delvigna-etal_2021} system by NEODyS\footnote{\url{https://newton.spacedys.com/neodys/NEOScan/}} and the internal monitoring system of the Minor Planet Center (MPC). To optimize and increase the chance to identify imminent impactors, these systems are designed to monitor objects when they are still on the NEO Confirmation Page (NEOCP), even before they get officially designated by the MPC. They have proved to be successful tools for the planetary defence community, after having correctly identified all of the most recent imminent impactors discovered by surveys since coming online, and having successfully issued alerts to astronomers for objects needing immediate confirmation and follow-up. 

Should a potentially hazardous object be found, the International Asteroid Warning Network (IAWN) is tasked with preparing an official impactor warning. This is only issued when the object is larger than 10 m in size and has an impact probability of greater than 1\% \cite{UNGA_78_72}. For objects not meeting this criteria, official warnings are not issued, but information is still disseminated to the public via ESA and NASA communication channels. In the context of imminent impactors, it is important to note that the vast majority of these objects are so small that they do not fall into the IAWN information regime, yet the institution stands ready should the situation arise.

This paper focuses on Meerkat, the imminent impactor monitoring system developed by ESA. 
The first version of the system became operational in 2021 \citep{fruhauf-etal_2021}. Since then, it has successfully contributed to the alert of all seven imminent impactors discovered. 

Meerkat is developed, operated, and maintained by the ESA NEO Coordination Centre\footnote{\url{https://neo.ssa.esa.int/}} (NEOCC). It serves as a rapid, lightweight complement to ESA's Aegis system \cite{fenucci-etal_2024, faggioli-etal_2023}, which uses more rigorous orbit determination methods, a deeper dynamical model and performs propagations on century-long timescales. The name was chosen to embody the exceptional vigilance and coordinated lookout behaviour of meerkats in the wild. Meerkats survive thanks to chosen guards who are constantly on the lookout for hazards (often from above, such as birds). The instant a threat is spotted, a sharp alarm call is sounded, and the whole colony reacts with remarkable speed and coordination. Much like the humble meerkat, the software keeps a watchful eye on the skies, swiftly issuing alerts for asteroid threats.

The original system was written in \texttt{python2} with a flight dynamics library that has since become deprecated. This motivated a significant software development plan to update the system to \texttt{python3} and use ESA's new flight dynamics library GODOT, a fully supported astrodynamics toolkit implemented in C++, with a Python API (GODOTPY) that exposes a subset of the C++ functionality for efficient algorithm execution. In this paper, we introduce the version 2.0 of Meerkat, which shows improved speeds, architecture and error rate, provide an overview of its current functionalities, and summarize its achievements during its operational lifetime.

\section{Orbit determination, impact scoring and Monte Carlo sampling}\label{sec:od_is_mcs}

\subsection{Systematic ranging}\label{subsec:sysranging}

Unconfirmed objects such as those in the NEOCP typically have short observational arcs, which may cause classical Gaussian orbit determination methods to fail \citep{milani-etal_2004}. 
From the observations, four of the six parameters necessary to determine an orbit are well known: the right ascension $\alpha$, declination $\delta$, the right ascension rate $\dot{\alpha}$ and declination rate $\dot{\delta}$. In order to constrain the orbital solution, systematic ranging explores a raster of the remaining poorly known parameters, selected as the topocentric range $\rho$ and topocentric range rate $\dot{\rho}$ following the approach in \cite{chesley_2004} and \cite{farnocchia-etal_2015c}. At each point of the raster, a fit is made to the observations to find the most likely orbital solutions. From this we can identify the correct region of sky for observational follow-up, allowing further observations to be taken, and traditional orbit determination methods to be performed. 

A given object will have a set of $n$ observations consisting of values $(\alpha_k, \delta_k)$ and possibly apparent magnitude $m_k$ at epochs $t_k,~ k=1, \dots, n$. The set of $n$ observations is composed of one or more \textit{tracklets}, short arcs of typically 3-4 observations from a given observatory. From this set of data, $\dot{\alpha}, \dot{\delta}$ can be estimated via a linear approximation using the first and last epochs $t_1$ and $t_n$. To perform systematic ranging, we assemble these in a so-called \textit{attributable}\cite{milani2005astrometry}, $\textbf{A} = (\alpha, \delta, \dot{\alpha}, \dot{\delta}) $ for a given epoch, which we select as the first epoch $t_1$. 

The raster is set up as a $N \times N$ grid in $(\rho, \dot{\rho})$ space, $\{ (\rho_{i},  \dot{\rho}_{j}) \ |\ i, j =1\ldots N\}$. At each grid point the orbit determination is performed using a least-squares method, following the procedure in \cite{milani-gronchi_2009}. The best-fit attributable for a given grid point $\mathbf{A}_{ij}(\rho_{i}, \dot{\rho}_j)$ is found by minimising the target function:
\begin{equation}
    Q = \mathbf{\xi}^T \mathbf{W} \mathbf{\xi}, 
    \label{eq:cost-function}
\end{equation}
where $\mathbf{\xi}$ is the observed minus computed astrometric residuals of the fit $\mathbf{\xi} = (\Delta\alpha_1,  \Delta\delta_1, \ldots, \Delta\alpha_n,\Delta\delta_n) $
and $\mathbf{W}$ is the $(2n \times 2n)$ weight matrix. The weight matrix is constructed by packing the inverted covariance matrix of each observation in a block diagonal: 

\begin{equation}
\mathbf{W} = 
\begin{bmatrix}
    \mathbf{C}^{-1}_{1} & 0 & \cdots & 0 \\
    0 &  \mathbf{C}^{-1}_{2}  & \cdots & 0 \\
    \vdots & \vdots & \ddots & \vdots \\
    0 & 0 & \cdots & \mathbf{C}^{-1}_{n}
\end{bmatrix}.
\end{equation}

The covariance matrix $\mathbf{C}_k, k=1, \dots, n$ of each observation has a contribution from astrometric uncertainty and one from timing uncertainty:

\begin{equation}
\mathbf{C}_{k} = 
\begin{bmatrix}
    \sigma_{\alpha_{k}}^2  & 0 \\
    0 & \sigma_{\delta_{k}}^2
\end{bmatrix} \ + \ \
\sigma^2_{t_{k}} \cdot
\begin{bmatrix}
    \dot{\alpha}_{k}^2& \dot{\alpha}_{k}\dot{\delta}_{k} \\
    \dot{\alpha}_{k}\dot{\delta}_{k} & \dot{\delta}_{k}^2
\end{bmatrix}, \quad k=1,\dots,n
\end{equation}
where $\sigma_{\alpha_{k}}$, $\sigma_{\delta_{k}}$ and $\sigma_{t_{k}}$ represent the uncertainty in the right ascension, declination and epoch, respectively. By default, $\sigma_{\alpha_{k}}$, $\sigma_{\delta_{k}}$ are selected using the weighting scheme from \cite{veres-etal_2017}, and the timing uncertainty is set to $\sigma_{t_{k}} = 1$~s. In addition, $\sigma_{\alpha_{k}}$ is scaled by $\cos(\delta_k)^{-1}$ to account for geometric scaling on the celestial sphere. 

The best-fit attributable is found using an iterative differential correction method with steps
\begin{equation}
    \label{eq:attributable-step}
    \Delta \mathbf{A}=-\left(\mathbf{B}^{T} \mathbf{W} \mathbf{B}\right)^{-1} \mathbf{B}^{T} \mathbf{W} \mathbf{\xi},
\end{equation}
where $\mathbf{B} = {\partial \mathbf{\xi} } / \partial \mathbf{A}$ is the $(2n \times 4)$ design matrix. The same initial attributable $\mathbf{A}^{\text{init}}$ is used as a starting guess for each point.
The astrometric weighted root mean square error (WRMSE) is computed for each grid point as
\begin{equation}
    \label{eq:wrmse}
    \mathrm{WRMSE} = \sqrt{\frac{1}{2 n} \mathbf{\xi}^{T} \mathbf{W} \mathbf{\xi}}.
\end{equation}
This provides a measure for the quality of the fit across the grid. The systematic ranging algorithm is initiated with the same initial attributable for each grid point and terminated upon reaching a change in WRMSE smaller than $10^{-2}$ or a maximum of four iterations. Each best-fit attributable is then converted to Keplerian orbital elements, $\textbf{Y}_{ij}$. For $\rho < 0.1$ au and $|\dot{\rho} | < 12 $ km s$^{-1}$, the geocentric orbital elements are also computed. 

% Posterior probability
The error function $f_{\text{err}}$ of normally distributed errors is combined with a prior probability density function $f_{\text{prior}}$ to compute a posterior probability density $f_{\text{post}}$ for the topocentric range and range rate as in \cite{farnocchia-etal_2015c}:
\begin{equation}
f_{\text {post }} \propto f_{\text {err }} f_{\text {prior }}, 
\end{equation}
\begin{equation}
f_{\text {err }}=\exp \left(-\frac{1}{2} \xi^{T} \mathbf{W} \xi \right), 
\end{equation}
\begin{equation}
f_{\text {prior }} = f_{\text {uni}} f_{\text{size}}(H)=\rho^{2-5 \eta}.    
\end{equation}

\begin{figure}[!ht]
    \centering
    \includegraphics[width=\linewidth]{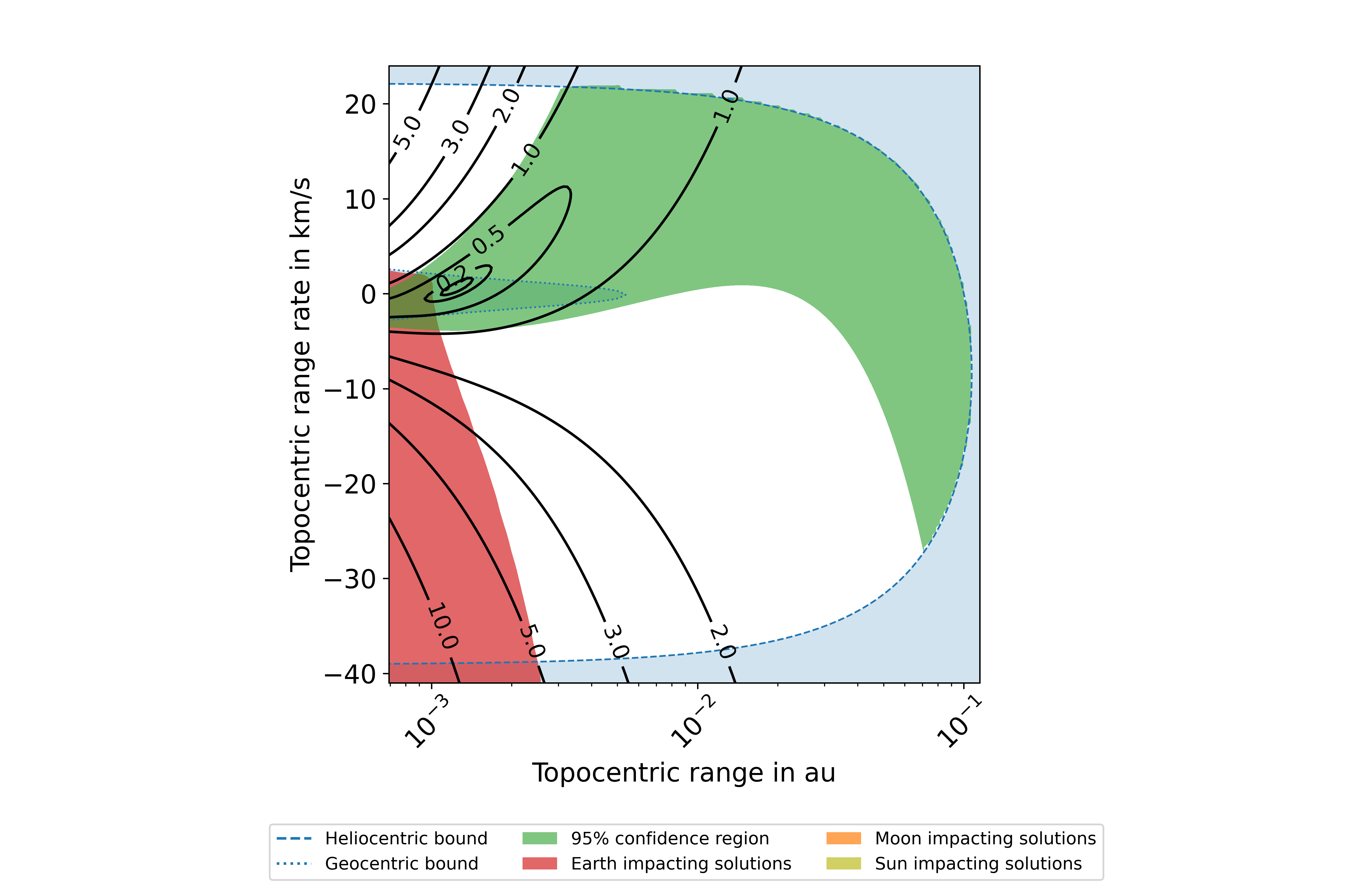}
    \caption{Systematic ranging plot for object P12cRbG, after 3 observations. The WRMSE, 95\% confidence region and Earth-impacting regions are indicated. In addition, the borders for unbound-heliocentric solutions and bound-geocentric solutions are shown as dotted and dashed blue lines respectively.}
    \label{fig:ranging-plot-P12cRbG-LHS}
\end{figure}

The choice of the prior is not trivial, and has signifiant effect on the output results. It has previously been found that a uniform distribution is good at identifying potential impactors and distinguishing scientifically interesting objects \cite{farnocchia-etal_2015c}. The prior $f_{\text{prior}}$ is based on a uniform distribution of objects in Cartesian position and velocity space  $f_{\text {uni}} \propto \rho^2 $, and a power-law size distribution, $f_{\text{size}}(H) \propto \rho^{-5 \eta}$, where for NEOs the exponent $\eta$ takes a value between 0.35 and 0.47 \citep{bowell-etal_1989, bottke-etal_2002, harris_2002, stuart-binzel_2004}. The value of $\eta = 0.41$ is selected, resulting in the $f_{\text {prior }}$ consistent with a uniform distribution. By default the grid uses a logarithmic scaling for $\rho$ for better sampling at small $\rho$ values. %In this instance, the prior probability has an additional factor of $\rho$ to account for the transformation.
We exclude any unbound heliocentric orbit solutions from our posterior distribution by defining all fitted grid points with eccentricity $e \geq 1$ the value of $f_{\text {post }} = 0$, restricting the posterior distribution to the so-called \textit{admissible region} \citep{milani-etal_2004}. A 95\% confidence region in $(\rho, \dot{\rho})$-space is attained by summing $f_{\text {post }}$ values from the minimum upwards until a 5\% boundary limit is found. An example of the systematic ranging output for NEOCP object P12cRbG is given in Fig. \ref{fig:ranging-plot-P12cRbG-LHS}. 

The grid of fitted absolute magnitude values $\mathbf{H}_{ij}$ is found 
using an iterative least squares in much the same way as the orbit determination, adapting Eqs.~\eqref{eq:cost-function}, \eqref{eq:attributable-step}, \eqref{eq:wrmse}. In this case we compute the best-fit absolute magnitude $H$ from the photometric data considering the previously found best-fit attributable. We use a weight matrix $\mathbf{W}$ composed of the apparent magnitude root mean square error $1 / \sigma^2_{mag}$ along the diagonal, the residuals $\mathbf{\xi}$ become the observed minus computed apparent magnitude residuals, and the design matrix becomes $\mathbf{B} = \partial\xi / \partial m$.  By default the apparent magnitude uncertainty is set to $\sigma_{mag}=0.2$ mag. The absolute magnitude $H$ is modelled with the $HG$ system \citep{bowell-etal_1989} using $G=0.15$. From the absolute magnitude, the object diameter is computed assuming an albedo value of 0.14, following \cite{bowell-etal_1989}, \cite{harris_2002}. 

By default, Meerkat will only process objects in the NEOCP with observations $3 \leq n \leq 30$. Beyond this computation time starts increasing, and it is assumed that the arc length and curvature will be significant enough for other orbit determination systems like Aegis \citep{fenucci-etal_2024} to process them. The grid is set up with side length $N=160$, and limits of $10^{-4}$ au to $10^{2}$ au in $\rho$ and $-$100 km s$^{-1}$ to $+$100 km s$^{-1}$ in $\dot{\rho}$. Before the orbit determination is begun, these limits are cropped to an approximation of the admissible region. The admissible region is estimated by using the initial attributable at points near each of the grid limits to find an $e \leq 1$ boundary. The lower $\rho$ limit is additionally cropped to have an absolute magnitude $H \leq 34.5$. 
After the orbit determination across the grid is complete, if the found 95\% confidence region is less than 2\% of the total grid size, the grid limits are cropped again and systematic ranging is performed in a so-called `zoomstep'. In each axis the new grid limits are found by taking the limits of the confidence region and adding a padding equal to $1.5\times$ the confidence region width. By re-performing systematic ranging on a denser grid in the proximity of the most likely solutions, the accuracy of the output results is increased. A maximum of one zoomstep is performed to ensure the algorithm remains fast. 

\subsection{Impact Scoring} 
\label{subsec: impact-scoring}
The best-fit orbit of each grid point is propagated forward 30 days from the latest observation epoch. Impacts with the Earth (defined as 50 km above the Earth equatorial radius, totalling $6.42814\times 10^{3}$ km from the geocentre\footnote{An impact was redefined from 0 km to 50 km above the Earth's surface from Meerkat v1.7 to v2.0.}) are registered if found. If not, the minimum close approach distance and time is found by optimisation with GODOT. The impact score is computed as the sum of posterior probabilities for all Earth-impacting grid points over the total posterior probability sum:

\begin{equation}
\label{eq:IP}
   \text{Impact score} = \frac{
        {\displaystyle \sum_{i=1}^{N} \sum_{j=1}^{N}
        f_{\text{post},ij} \cdot \kappa_{ij}}}
        {{\displaystyle \sum_{i=1}^{N} \sum_{j=1}^{N}
        f_{\text{post},ij} }
    }
\end{equation}

\begin{equation}
\label{eq:impact-flag}
\kappa_{ij} =
\begin{cases}
1 & \text{if point} \ (\rho_i, \dot{\rho}_j) \ \text{is an impactor,} \\
0 & \text{otherwise.}
\end{cases}
\end{equation}

Impacts with the lunar surface, defined as $1.73814\times10^3$ km from the lunar centre, are also found by propagation. A solar impact is computed by approximation, simply by checking if the object perihelion $q$ is less than or equal to the solar radius $R_{\odot} = 6.96340 \times 10^{5}$ km. Both lunar and solar impacting solutions are indicated on the systematic ranging plot (see Figs. \ref{fig:ranging-plot-P12cRbG-LHS} and \ref{fig:ranging-plot-full-2024XA1}) with orange and yellow shading respectively, if found. The terrestial impact (or close approach) time and position in Cartesian ICRF coordinates is packaged for each grid point in $\mathbf{K}_{ij}$.

In a similar fashion to Eq.~\eqref{eq:IP}, we compute a weighted sum posterior probability score to estimate several other object characteristics, such as the possible different orbital classes and estimated size. A list of output scores is given in Section  \ref{subsec:meerkat-dashboard}. Note that the terms \textit{score} and \textit{probability} are often used interchangeably here, despite the fact Eq.~\eqref{eq:IP} is computed over a fine grid and so is not analytical. 

\subsection{Interstellar Object Scoring}
While Meerkat is conceived to handle Solar System objects, it is scientifically interesting to promptly notify interstellar objects. To this purpose, an interstellar object (ISO) score is computed with an n-sigma t-test. The test compares a hypothetical population with mean eccentricity $e=1$ and 0 standard deviation with the population in the bound region of the grid, with weighted mean eccentricity $\bar{e}$ and standard deviation $\sigma_{e}$: 
\begin{equation}
\label{eq:iso-score}
\text{ISO score} = 
w_{\text{ISO}} \cdot \left( \frac{\sqrt{N_e} \ (1 - \bar{e})}{n_{\text{ISO}} \ \sigma_e} \right) ^{-1}.
\end{equation}
In Eq.~\eqref{eq:iso-score}, $N_e$ is the number of grid points with $e < 1$, and $n_{\text{ISO}}$ is a user-defined parameter indicating the sigma level to test, nominally $n_{\text{ISO}}=2$. The factor $w_{\text{ISO}}$ is a weighting function to scale the score with the number $n$ of observations: 
\begin{equation}
\label{eq:iso-weight}
w_{\text{ISO}} = 
\frac{n^3}{\left( 16 + \left| n - 16 \right| \right)^3}.
\end{equation}
An ISO score $> 1$ indicates the object having eccentricity $e=1$ is within the confidence region and there are enough observations for the result to be reliable. A typical object in the NEOCP will have an ISO score of $\sim10^{-5}$, as such an ISO score of 0.1 is chosen as a threshold to start sending alerts for a potential ISO. With this threshold, an ISO alert would have been sent for 1I/`Oumuamua \citep{meech-etal_2017} at 16 observations, and 3I/ATLAS \citep{seligman-etal_2025} at 15 observations. 2I/Borisov would have been different, as it took many more observations to be confirmed as an ISO. When it was in the NEOCP, it was only starting to be considered as a possible ISO after $\sim$9 days and 70 observations \citep{guzik-etal_2020}, which is outside our observation number threshold of $n\leq30$, so would have been run manually by NEOCC analysts. 

\subsection{Dynamical model}
One of the main changes in the new implementation of Meerkat is in the libraries used for the dynamics of the Solar System. From the Meerkat v2.0 onwards, all dynamical computations are performed using GODOT\footnote{\url{https://godot.io.esa.int/}}, an astrodynamics library developed by \cite{godot-software}. GODOT is a comprehensive and highly versatile library, allowing users to develop solutions to a wide variety of space flight dynamics problems. Propagations are performed using the \texttt{ballistic propagator} class, with an Adams stepper, tolerance of $10^{-9}$ and max-steps of $10^7$. Impacts are found using an \texttt{event trigger} class of tolerance $10^{-6}$~s. 
A simple dynamical model which includes the gravitational attraction of the Sun, the Earth, and the Moon is employed for all propagations. The positions of the celestial bodies are retrieved from the JPL Planetary and Lunar Ephemeris DE440 \citep{park-etal_2021} together with their associated gravitational parameters. 

\subsection{Monte Carlo Sampling}\label{subsec:mcsampling}
To compute other information about the object, such as possible impact locations and detection probabilities, we additionally perform a Monte Carlo analysis. After systematic ranging, 1000 Monte Carlo sample orbits are generated (indexed over $s$) following the steps in \cite{farnocchia-etal_2015c}:
\begin{enumerate}
    \item Randomly draw a topocentric range and range rate value $\rho_s \sim  \mathcal{U}(\rho^{\min}, \rho^{\max}) $ ; $\dot{\rho_s} \sim  \mathcal{U}(\dot{\rho}^{\min}, \dot{\rho}^{\max}) $;
    \item Randomly draw a posterior probability density value  $\gamma_s \sim  \mathcal{U}(0, f_{\text {post }}^{\max}) $;
    \item Compute the best-fit attributable $\mathbf{A}_{s}(\rho_{s}, \dot{\rho}_s)$ and posterior probability density $f_{\text {post},s} \ (\rho_{s}, \dot{\rho}_s)$;
    \item Keep this sample only if  $f_{\text {post},s} \geq \gamma_s $, otherwise restart from step (1);
    \item Add Gaussian noise to the fitted attributable, defined by the covariance of the fit $
    \mathbf{C}_s=\left(B^{T}_s \mathbf{W} B_s\right)^{-1}$ to  $\mathbf{A}_{s}\left(\rho_{s},  \dot{\rho}_{s}\right)$.     
\end{enumerate}

The notation $\mathcal{U}(x_1, x_2)$ refers to a uniform distribution between $x_1$ and $x_2$. These samples are propagated forward 30 days from the latest observation epoch. Impacts with the Earth and Moon as well as close approaches to Earth are registered if found as in Section~\ref{subsec: impact-scoring}. From this process a sample of best-fit attributables $\mathbf{A}_{s}^{\text{MC}}$, Keplerian orbital elements $\textbf{Y}_{s}^{\text{MC}}$ and impact (or close approach) locations and times $\mathbf{K}_{s}^{\text{MC}}$ is produced. Found impact times and locations are plotted to better visualize the possible impact locations (see Section \ref{subsec:other-plots}). 

\subsection{Detection Probability for Follow-up Pointing}\label{subsec:detprob}
Meerkat computes ephemerides for all propagated Monte Carlo samples 5 days from the epoch of last observation with a time-step of 30 minutes. For each sample, several helpful parameters are computed including the position in the plane of sky, apparent magnitude, angular velocity, lunar separation, solar elongation and position angle. This is computed for a list of specified MPC stations as well as from the geocentre. 
For each time-step, the best telescope pointing is computed as the median $\alpha$ and $\delta$ considering only samples that are brighter than the limiting magnitude of the station and above the station horizon limit at the time. The best pointing from the geocentre is an exception as it is determined assuming an unbounded limiting magnitude and no horizon. It was found that the median statistic produces better telescope pointing than the mean, due to the mean being susceptible to large outliers which can be produced from systematic ranging and subsequent sampling \citep{fruhauf-etal_2019}. 

The detection probability from the geocentre at each time-step is also computed for a range of field-of-view (FoVs) from 5' $\times$ 5' to 500' $\times$ 500'. This is computed by counting the fraction of samples that are within a given FoV diameter centered at the median best pointing $\alpha$ and $\delta$.  As the detection probability is computed from the geocentre, it does not take into account a local horizon or limiting magnitude. Instead, the station-specific observability is shown superimposed on the detection probability contours, as shown in Fig. ~\ref{fig:station-selector} (see Sec \ref{subsec:other-plots}).  

\subsection{Extra parameters for likely impactors}\label{subsec:extraparams}
Given an impact score of over 60\%, extra statistics are computed relevant to imminent impactors. Using the impacting Monte Carlo samples, the minimum, maximum and median impact velocity and impact flight-path angle (FPA), defined as the angle between the velocity vector and the normal to the position vector, are computed. This is useful to ascertain if a potential fireball will be observable from a given location. 

In addition, the time the object will enter the umbra of the Earth's shadow is computed for the samples with the earliest, latest and median impacting times. This gives observers an estimated range for the observation cut-off time, so they can observe right up to the final minutes before impact. 

\section{Meerkat architecture}
Meerkat implements the methods described above using an automated data pipeline for NEOCP scanning. 
The NEOCP list is downloaded in the form of the \texttt{neocp\_obs} table from the MPC PostgreSQL\footnote{\url{https://data.minorplanetcenter.net/mpcops/documentation/replicated-tables-schema/}} database. Meerkat checks the NEOCP approximately every 45 seconds to see if there are any new observations added or altered. 

If found, objects with updates are passed into the analysis loop. All observations are processed to ensure correct formatting and spot obvious errors, such as duplicated tracklets. The objects are then queued according to their last computed impact score, following the  priority scheme in Table \ref{tab:priority-comp}. This is used to ensure that potentially impacting objects are computed with upmost importance, and do not have to wait for other objects to finish being analysed. 

\begin{table}[!ht]
    \centering
\begin{tabular}{c c}
\hline
\textbf{Priority} & \textbf{Impact Score} \\
\hline
1 & $\geq 10\%$ \\
2 & $\geq 1\%$ \\
3 & N/A (\textit{New object}) \\
4 & $< 1\%$ \\
\hline
\end{tabular}
    \caption{The Meerkat priority scheme for object computation. A value of 1 is the highest priority.}
\label{tab:priority-comp}
\end{table}

\begin{figure}[!ht!]
    \centering
    \includegraphics[width=0.5\textwidth]{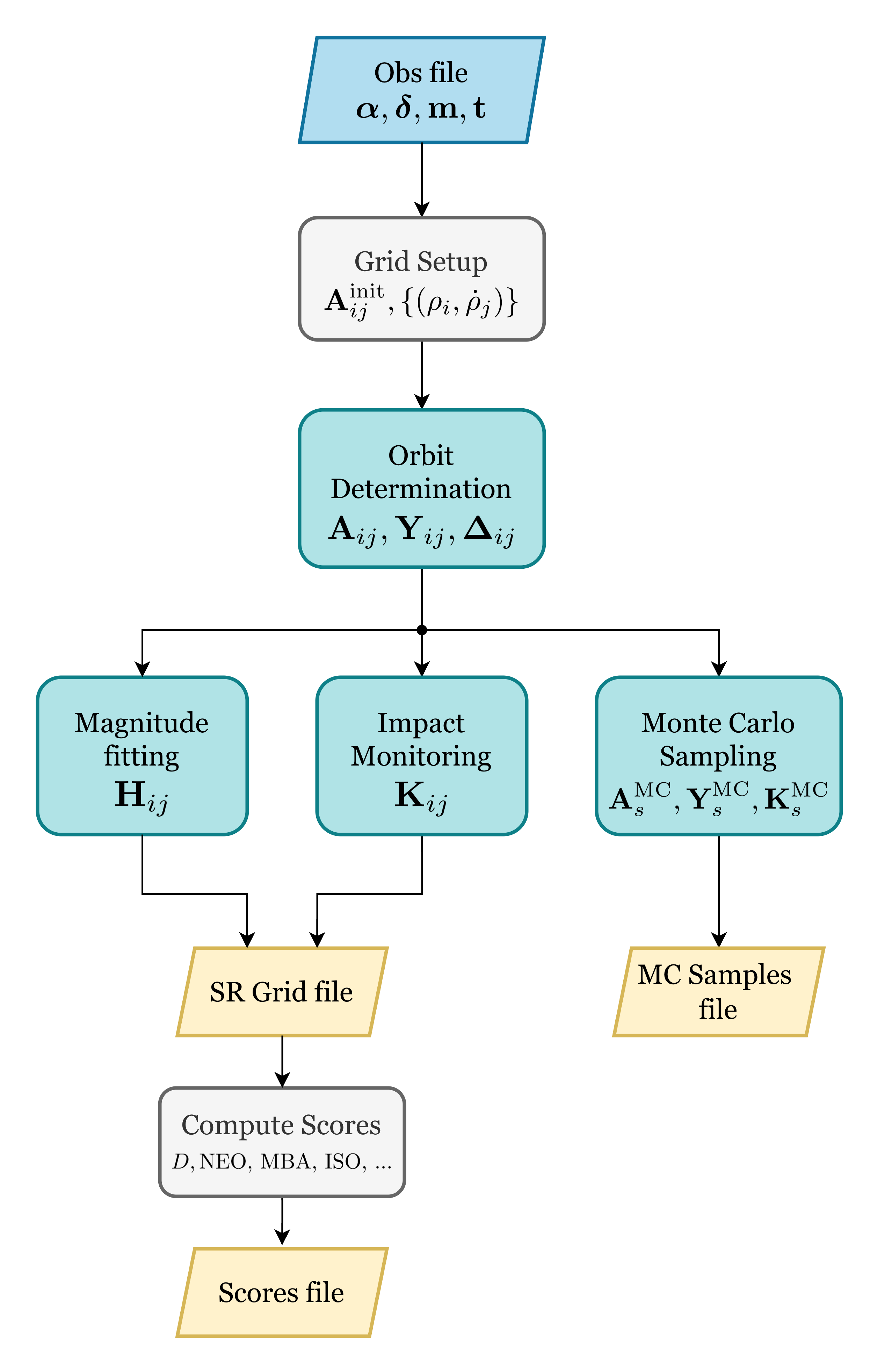}
    \caption{Schematic of the Meerkat systematic ranging module. An initial attributable $\mathbf{A}_{ij}^{\text{init}}$ is fitted recursively to the input observations for points of fixed topocentric range and range-rate values $\{ (\rho_{i},  \dot{\rho}_{j}) \}$. The grid of best-fit attributables $\mathbf{A}_{ij}$, along with their likelihood $\mathbf{\Delta}_{ij}$ and Keplerian elements $\mathbf{Y}_{ij}$ are then used to compute absolute magnitudes $\mathbf{H}_{ij}$ and impact (or close approach) locations and times $\mathbf{K}_{ij}$. Several scores are computed, estimating the object size and orbital class. Monte Carlo samples are also drawn. Three files are outputted as indicated in yellow, while teal boxes indicate stages where multiprocessing is used.}
    \label{fig:sysranging_fc}
\end{figure}

Each object is passed sequentially to the systematic ranging module, shown schematically in Fig. \ref{fig:sysranging_fc}. This uses the steps in Sections \ref{subsec:sysranging}-\ref{subsec:mcsampling} and \ref{subsec:extraparams} to produce three output files: the \texttt{SR Grid} file, the \texttt{Scores} file and the \texttt{MC Samples} file. The \texttt{SR Grid} file contains the fitted attributables, orbital elements and absolute magnitudes for each grid point $\mathbf{A}_{ij}, \textbf{Y}_{ij}, \mathbf{H}_{ij}$, along with the computed impact (or close approach) times and locations for each grid point $\mathbf{K}_{ij}$. The derived WRMSE and posterior probability values, stored together in $\mathbf{\Delta}_{ij}$, are also outputted in this file. The \texttt{Scores} file contains the size and orbital class scores computed from the systematic ranging grid (see Section \ref{subsec:meerkat-dashboard}.). Finally, the \texttt{MC Samples} file contains the drawn Monte Carlo samples as well as any extra computed parameters for likely impactors, following Sections \ref{subsec:mcsampling} and \ref{subsec:extraparams}. 

\begin{figure}[!ht!]
    \centering
    \includegraphics[width=0.75\textwidth]{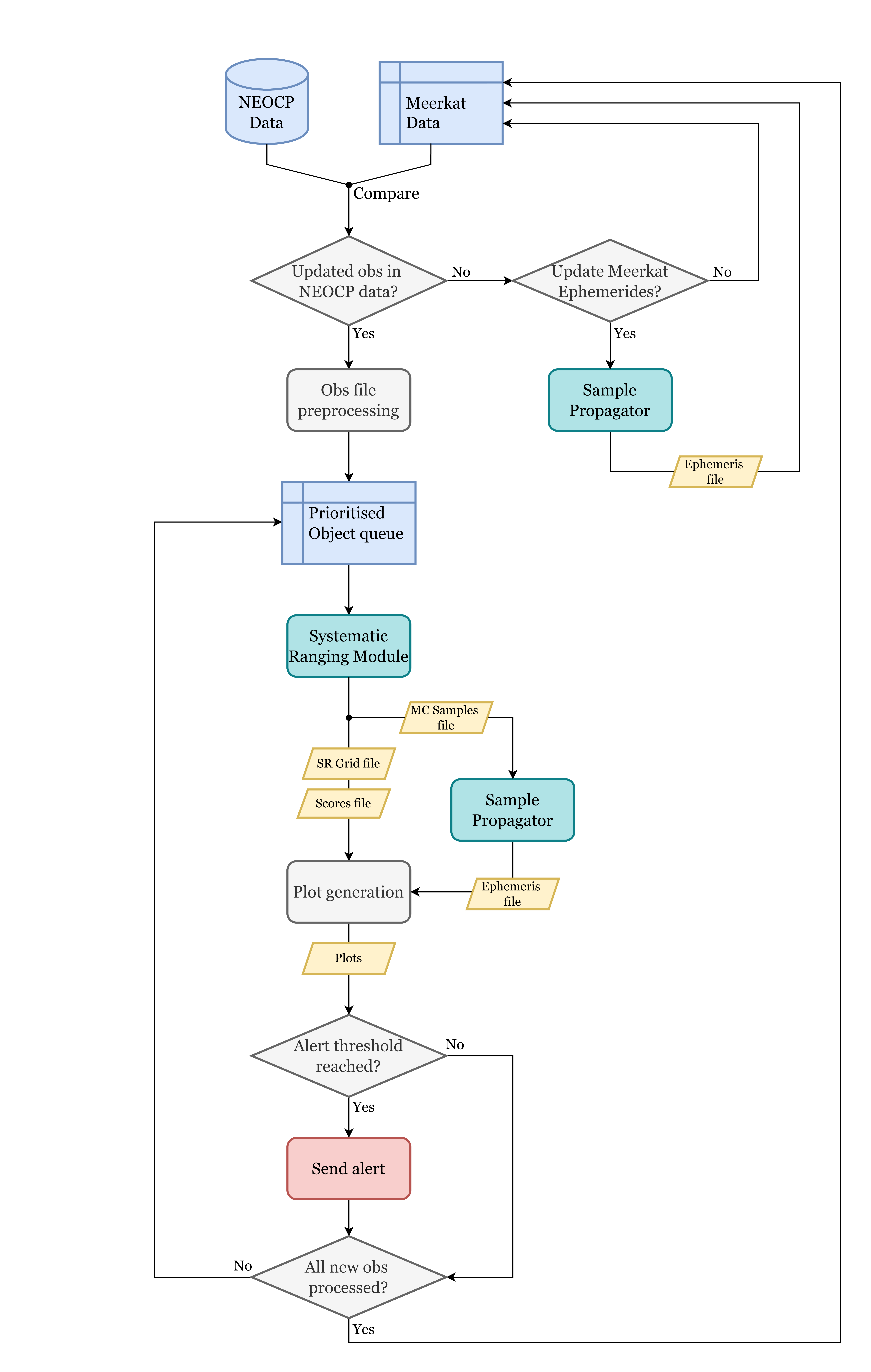}
    \caption{Schematic of the Meerkat data pipeline. The NEOCP is continually checked for new or updated observations. Once found, observations are processed and their objects put in a prioritised queue based on their last computed impact scores. Each object in the queue is analysed sequentially using the \texttt{systematic ranging module} (Fig. \ref{fig:sysranging_fc}), with the outputs used to compute a best pointing and ephemeris via the \texttt{sample propagator} (Section \ref{subsec:detprob}) and generate informative plots (Sections \ref{subsec:meerkat-dashboard} and \ref{subsec:other-plots}). If a given object meets the threshold criteria, an alert is issued to Meerkat subscribers (Section \ref{sec:meerkat-alerts}). If no updated observations are found, object ephemerides may be updated if older than two days. Boxes in teal represent where multiprocessing is used.}
    \label{fig:backend_fc}
\end{figure}

Following systematic ranging, the \texttt{MC Samples} file is passed to the \texttt{sample propagator}, which produces an ephemeris and best pointing for follow-up observations, as in Section \ref{subsec:detprob}. The \texttt{SR Grid, Scores, MC Samples} and \texttt{Ephemeris} files are passed to the plotting scripts to produce the plots described in Sections \ref{subsec:meerkat-dashboard} and \ref{subsec:other-plots}. If a given object meets the threshold for an alert, an email is sent to Meerkat subscribers, following Section~\ref{sec:meerkat-alerts}. This cycle continues until the prioritised object queue is empty. If no observation updates are found in a given NEOCP check, Meerkat will update the ephemerides of it's currently stored objects, given the ephemeris is more than two days old. 
This data pipeline is summarised in Fig. \ref{fig:backend_fc}.
Note that the MPC PostgreSQL database as input data source has been introduced in the new Meerkat 2.0 version, providing a more robust and faster access to NEOCP data.

Meerkat is completely written in Python. Meerkat v1.7 still had dependencies on \texttt{python2.7}, which has been discontinued in 2020 and therefore no longer suitable for an operational software. These dependencies were removed in the new Meerkat v2.0, which now runs on a modern \texttt{python3.11}. 
The software is maintained under version control using a repository hosted on GitLab\footnote{\url{https://about.gitlab.com/}}.
Continuous Integration and Continuous Deployment (CI/CD) pipelines are employed within GitLab to automate testing, Docker image creation, and deployment. Meerkat runs using Docker images generated during the CI/CD pipeline, each containing fixed versions of Python and all required packages and modules. This setup ensures that the software remains stable and unaffected by future Python or dependency upgrades.
The Meerkat data pipeline summarized in Fig.~\ref{fig:backend_fc} runs through a fully dockerized service\footnote{\url{https://www.docker.com/}}, with containers orchestrated via Docker Swarm to ensure scalability, portability, and efficient resource management.
Meerkat employs multiprocessing for several processes, exploiting the parallel nature of many of the computations. In the orbit determination, magnitude fitting and impact monitoring, each column of the grid is assigned a process for multiprocessing. In the Monte Carlo sampling, each sample is assigned a process to perform the rejection sampling, as well as subsequent magnitude fitting, impact monitoring and follow-up pointing determination. 
The deployment is carried out on a 32-core virtual machine hosted on the ESA Cloud infrastructure, using all 32 cores for multiprocessing steps. The use of multiprocessing allows Meerkat to be scalable; as the rate of detections steadily increases into the future, much of the required computational speed increase can be met by increasing the number of resources available.

Ideally alerts should be issued in minutes to give astronomers time to perform follow-up observations before the object is lost. Since Meerkat 2.0 has become operational, the average time to issue an alert is 56 s, with all objects processed without errors taking under 3.5 minutes. This is an increase in computational speed of approximately 10\% from the previous version. 
The choice of 45 seconds between checks for NEOCP updates is somewhat arbitrary, but is currently sufficient to deliver timely alerts from observation updates. Even with this high sampling rate, Meerkat spends on average 20.5 hours asleep each day, waiting for new NEOCP data if not updating ephemerides of existing objects. As the number of objects in the NEOCP steadily increases, the sampling parameter can be adjusted to meet increasing demand.

Meerkat has been designed to operate with high stability and robustness, and it has proven to be very reliable in continuous operations. Nevertheless, several mechanisms have been implemented to ensure proper functioning and redundancy. Two identical instances of Meerkat are running — OPE and PREOPE. If the OPE machine experiences any issue, operations can be quickly switched to PREOPE. Moreover, several scripts continuously check the operational status of Meerkat: a script verifies that all Meerkat Docker services are running correctly; another one compares the number of NEOCP objects in the Meerkat system with those in the Scout system. Both scripts send alerts via email and phone messages to NEOCC operators, allowing them to promptly investigate and resolve any issue. In addition, redundancy is further supported by the existence of independent systems worldwide that monitor imminent impactors, such as Scout and NEOScan.

\section{Meerkat alerts}
\label{sec:meerkat-alerts}

\subsection{Alert messages and service subscription}
Meerkat alerts are issued internally to the NEOCC team by email, text messages and direct phone calls for certain cases. Alerts are also announced to the astronomical community through a mailing list maintained by the ESA NEOCC. The Meerkat mailing list is open only to individuals directly involved in the NEO field, particularly to observers and researchers who promptly need access to potential impactor information for performing follow-up observations or other relevant work. To subscribe to the Meerkat mailing list, users may need to provide additional information, such as: 1) an affiliation and, if applicable, the associated observatory and observatory code; 2) a justification for the request and intended use cases; 3) if applicable, a selection of at least two MPECs reported by the requester in the last 10 years.

We stress that all the information accessible via the Meerkat mailing list relates to unconfirmed objects. Thus, all the data provided in the alerts is available for strict personal use, and it is not to be distributed further via any medium. Requesting users are also asked to agree to these usage conditions. Requesters not compliant with the requirements are redirected to the NEOCC web portal, where a dedicated page about imminent impactors\footnote{\url{https://neo.ssa.esa.int/imminent-impactors}} is now operational.

For external Meerkat subscribers, an alert will be issued if the impact score is greater than 10\% or the weighted mean distance of close approach (DCA) is less than 7 Earth radii. Internal to the NEOCC, alerts may be sent for more specific criteria, such as an ISO score greater than 0.1. 

All alerts have a \textit{New} or \textit{Update} label in the header to indicate if the object is new in the NEOCP list. If follow-up observations mean an object no-longer meets the alert criteria, subscribers are notified by an \textit{All-clear} email. Similarly, subscribers are notified of objects being removed from the NEOCP via an \textit{Object removed} email. In this way, it is easy to track the progress of an object's scores through successive Meerkat computations each time new observations are submitted.

\subsection{Meerkat dashboard}
\label{subsec:meerkat-dashboard}
\begin{figure}[!ht]
    \centering
    \includegraphics[width=\linewidth]{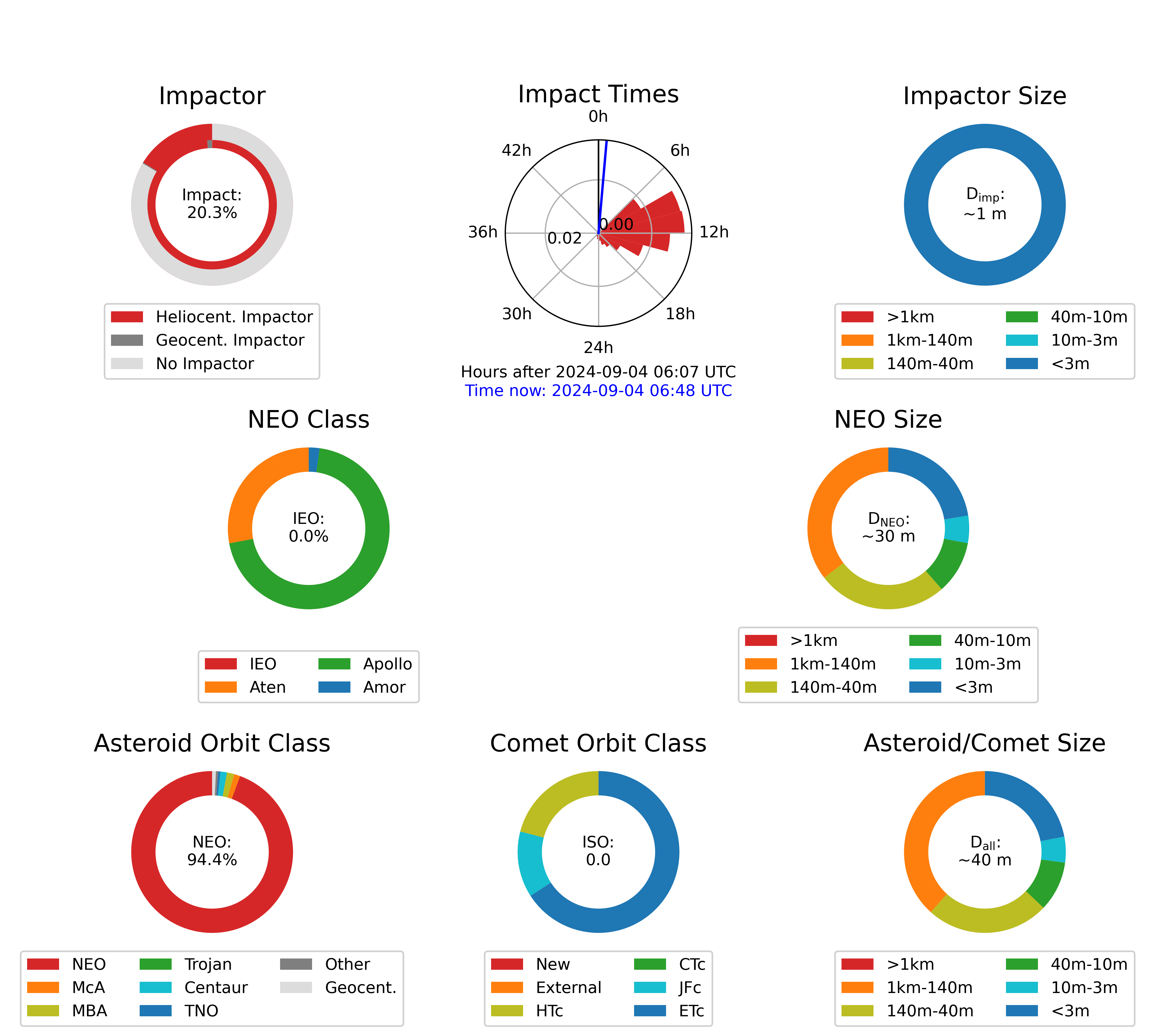}
    \caption{Dashboard of a Meerkat alert for NEOCP object CAQTDL2 after 4 observations (later designated 2024~RW1).}
    \label{fig:Dashboard_CAQTDL2}
\end{figure}

The first component of a Meerkat alert is a visual dashboard, summarising the possibility of impact and estimated object characteristics, as shown in Fig. ~\ref{fig:Dashboard_CAQTDL2}. The top of the dashboard indicates the NEOCP object name, the number of observations used in the analysis, and the corresponding arc length. 
Several pie charts are shown, grouped into rows: the top row shows information about Earth-impacting solutions, the middle row shows information about NEO solutions, and the bottom row considers all solutions. The individual pie charts are described below, where (a) to (c) indicates moving left to right: 

\vspace{1em}

\noindent Row 1, Earth-impacting solutions:
\vspace{0.3em}
\begin{enumerate}
  \renewcommand{\theenumi}{\alph{enumi}}
  \renewcommand{\labelenumi}{(\theenumi)}
  \item The impact score is shown as the thick outer ring, indicating the fraction of heliocentric impacting solutions (in red), the fraction of geocentric impacting solutions (in dark gray), and the non-impacting solutions (in light gray). The inner ring shows only the relative fractions of impacting solutions. The impact score is also given as a percentage inside the ring.
  \item The impact score as a function of the impact time, in bins of 2 hours and over a timespan of 48 hours from the last reported observation. The current epoch is indicated in blue.
  \item The estimated size distribution of the impacting solutions in different bin sizes, with the mean diameter value indicated in the centre of the ring. 
\end{enumerate}

\vspace{1em}

\noindent Row 2, NEO solutions:
\vspace{0.3em}
\begin{enumerate}
  \renewcommand{\theenumi}{\alph{enumi}}
  \renewcommand{\labelenumi}{(\theenumi)}
  \item The fraction of solutions corresponding to Inner Earth Orbit (IEO), Amors, Atens, and Apollos.
  \item The size distribution of NEO solutions, divided in different diameter bins. 
\end{enumerate}

\vspace{1em}

\noindent Row 3, All solutions:
\vspace{0.3em}
\begin{enumerate}
  \renewcommand{\theenumi}{\alph{enumi}}
  \renewcommand{\labelenumi}{(\theenumi)}
  \item The fraction of possible different orbital classes, including: NEO, Mars-crossing asteroid (MCA), main-belt asteroid (MBA), Jupiter Trojan, Centaur, trans-Neptunian object (TNO), and geocentric.
  \item The fraction of different types of cometary solutions following the definitions in \cite{levison-etal_1996}: Jupiter Family Comet (JFC), Encke Type Comet (ETC), Halley Type Comet (HTC), External Comet (EC) and New Comet (NC). The ISO score is presented in the centre of the pie chart.
  \item The size distribution, taking into account both asteroidal and cometary solutions. 
\end{enumerate}

\vspace{1em}

Separating the results in this way allows for more targeted analysis of different impact or close approach scenarios. For example, impacting solutions from systematic ranging typically have large absolute magnitude and hence smaller sizes than other NEO solutions, so it is important to distinguish the two cases when preparing a mitigation response.

Fig. ~\ref{fig:Dashboard_CAQTDL2} shows an example dashboard computed by Meerkat for the NEOCP object CAQTDL2. The computations were made with a tracklet of 4 observations, covering an arc of 0.40 h. The impact score was significant at 20.3\%, as indicated in the first pie chart of the first row, while the average diameter for the impacting solutions was only 1 m. From the second row, it can be understood that NEO solutions are most likely of Apollo type, while there is about 25\% possibility for Aten type, and a small probability for Amor class. The weighted mean diameter for the NEO solution was 30 m, although solutions with a larger or smaller diameter were also possible. In the last row, it can be understood that the majority of solutions obtained by the systematic ranging were of NEO type, about 94\%. There were also different types of cometary solutions, but not very likely. This object was later designated 2024 RW1 by the MPC and impacted the Earth on 4 September 2024 (see Section \ref{subsubsec:2024RW1}). 

\subsection{Informative plots}
\label{subsec:other-plots}
\begin{figure}[!ht!]
    \centering
    \includegraphics[width=0.85\textwidth]{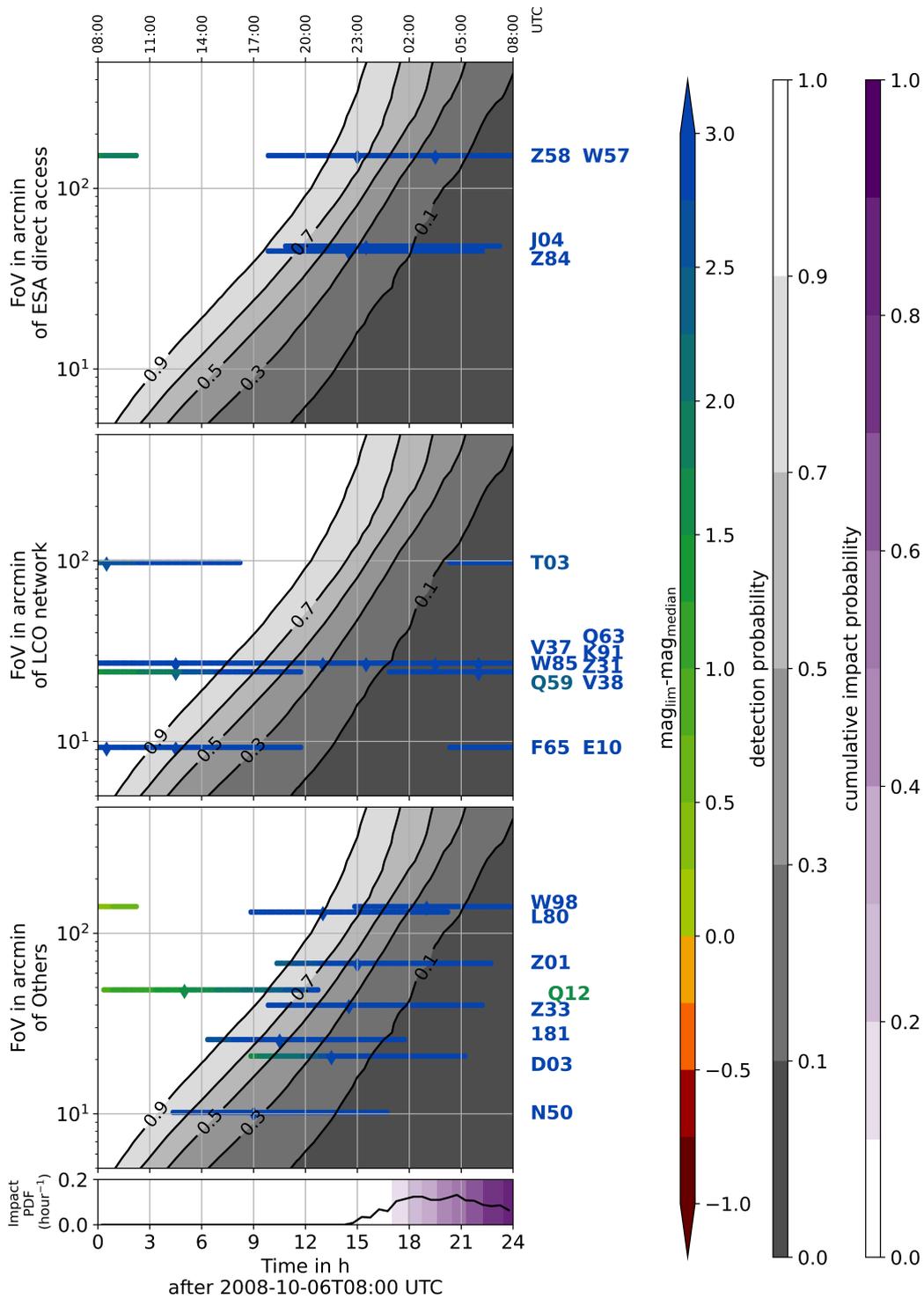}
    \caption{Station selector plot for the object K08T03C (later designed 2008 TC3) after 8 observations. The full station selector plot includes another column of contour plots to show the observability for several MPC stations.}
    \label{fig:station-selector}
\end{figure}

Along with the dashboard, Meerkat alerts include several other informative plots on the object analysed.
Fig.~\ref{fig:station-selector} shows a station selector plot. This is designed to help observers understand whether an object can be imaged with a telescope of a given field of view (FoV). In the plane of FoV vs. observation time, the plot shows the contour lines of the probability of detecting the object. The detection probability values are geocentric, rather than station dependent, hence an observer should independently analyse the visibility conditions from their site. 

For Meerkat internal users (ESA PDO personnel), the plot includes various MPC stations as horizontal bars. The length of the bar corresponds to the times the median pointing centre is above the local horizon and the colour corresponds to the difference between the station-dependent limiting magnitude and time-dependent sample median magnitude. The time the median pointing centre is at its zenith is given as a diamond. 

The impact probability density function (PDF) per hour is also shown on a panel at the bottom of the figure. Shaded behind the PDF in purple is the cumulative impact probability as a function of time.  

\begin{figure}[!ht]
    \centering
    \includegraphics[width=\linewidth]{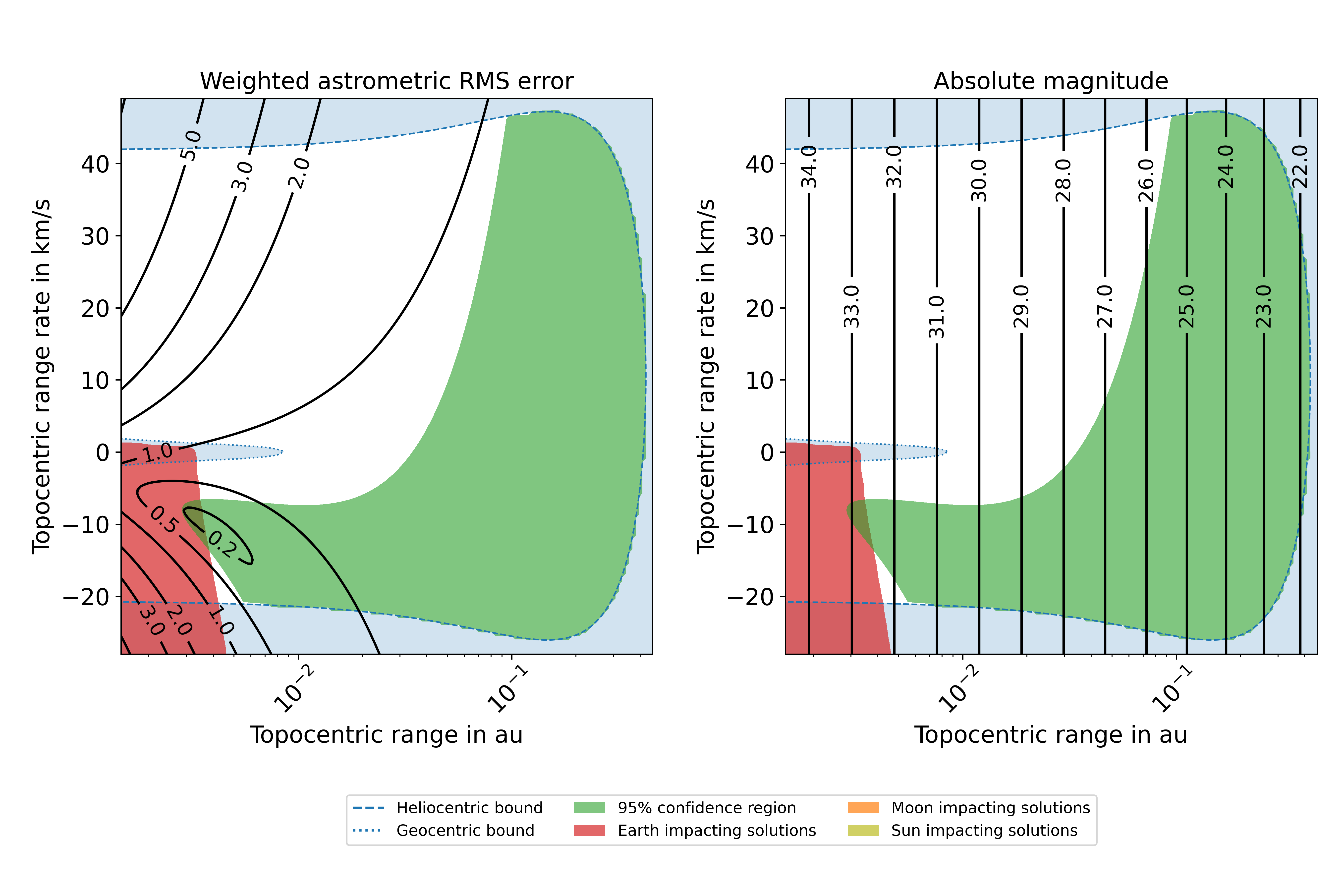}
    \caption{Systematic ranging plot for object C0WEPC (later designated 2024~XA1) after 4 observations. The WRMSE contours are shown on the left, while the contour lines of the absolute magnitude fit are shown on the right.}
    \label{fig:ranging-plot-full-2024XA1}
\end{figure}

The results of the systematic ranging are also shown in the Meerkat alerts, such as in Fig.  \ref{fig:ranging-plot-full-2024XA1}. Each panel is plotted in the plane of topocentric range rate $\dot{\rho}$ vs. topocentric range $\rho$. The boundaries of the heliocentric-bound solutions (dashed outer blue border) and geocentric-bound solutions (dotted inner blue spike) are indicated, as defined in \citet{milani-etal_2004}. The left panel shows the contour lines of the WRMSE from Eq.~\eqref{eq:wrmse}, while the right panel shows the contour lines of the absolute magnitude fit. In addition, both panels show the 95\% confidence region of the orbital solution in green shading, along with the Earth, Moon, and Sun impacting solutions in red, orange, and yellow shading respectively. These plots help the user to understand the orbital solutions compatible with the observations, the size of the uncertainty, and to see how likely impacting solutions are compared to the whole uncertainty region. 

\begin{figure}[!ht]
    \centering
    \includegraphics[width=\linewidth]{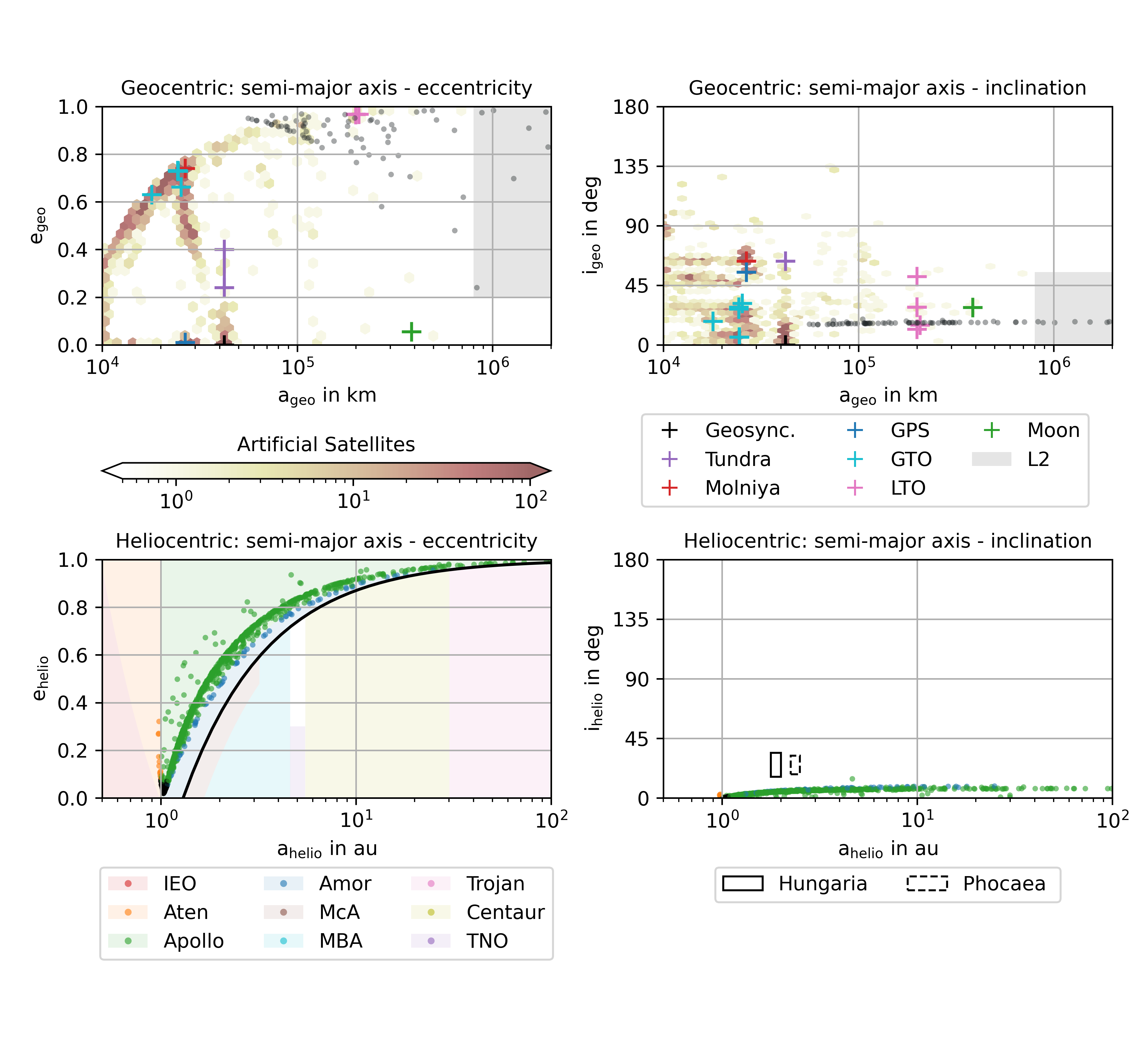}
    \caption{Scatter plots of object P12cRbG after 3 observations, indicating the distribution of heliocentric and geocentric orbital elements of the Monte Carlo samples.}
    \label{fig:scatter-plot}
\end{figure}

Additional scatter plots are presented to better understand the nature of the object, as shown in Fig. \ref{fig:scatter-plot}. First, a pair of scatter plots in the planes of geocentric eccentricity vs. geocentric semi-major axis and geocentric inclination vs. geocentric semi-major axis are shown. Both plots show the density of artificial satellites as a function of the geocentric orbital elements. Some particular regimes are highlighted with different colours, including: geosynchronous satellites, Tundra satellites, Molniya orbits, global position system (GPS) satellites, lunar transfer orbits (LTO), and geosynchronous transfer orbits (GTO). Artificial satellite data is obtained from the CelesTrak satellite catalogue (SATCAT) \citep{celestrak-satcat}. Any Monte Carlo samples drawn corresponding to geocentric orbital solutions are plotted as grey dots over the histograms to indicate the possibility of the object being artificial.

Another pair of scatter plots in the plane of heliocentric eccentricity $e$ vs. heliocentric semi-major axis $a$ and in heliocentric inclination $i$ vs. heliocentric semi-major axis $a$ are also presented. The panel in $(a,e)$ highlights the boundary of the NEO region at perihelion distance of 1.3 au, while different orbital regimes are indicated by shaded areas. On the other hand, the panel in $(a,i)$ highlights the Hungaria and the Phocaea regions, which are main-belt regions densely populated by the presence of asteroid families, and isolated from the rest of the main belt because of their relatively high inclination. Both panels show the distribution of Monte Carlo orbital solutions with the circle colour indicating the sample orbital class. This plot helps users to understand the distribution of orbital solutions with an indication of possible natural origin. 

\begin{figure}[!ht]
    \centering
    \includegraphics[width=\linewidth]{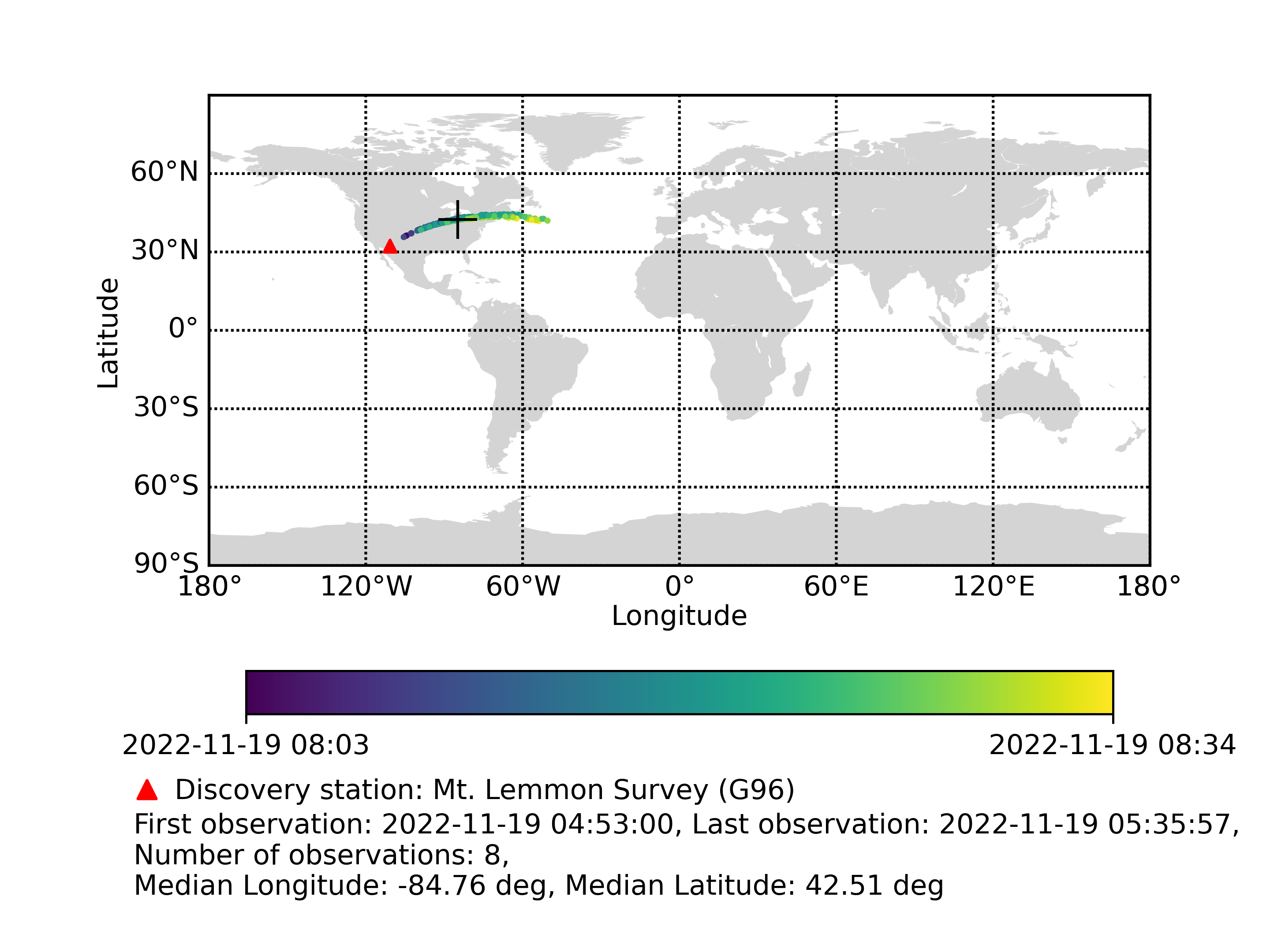}
    \caption{Impact plot for object C8FF042 after 8 observations, later designated 2022 WJ1.}
    \label{fig:impact-plot}
\end{figure}

Finally the Meerkat alert contains a map showing the impact corridor obtained by the Monte Carlo sampling. The median impact location in longitude and latitude is marked with a plus symbol, and range of impact times given as a colourmap. The discovery station is also highlighted with a red triangle mark. An example of an impact plot for object C8FF042 is shown in Fig.  \ref{fig:impact-plot}. This object was later designated 2022 WJ1, after impacting near the Great Lakes region in North America. Further information on 2022~WJ1 is given in Section \ref{subsubsec:2022WJ1}

\section{Successful Meerkat alerts}
\begin{figure}[!ht]
    \centering
    \includegraphics[width=\linewidth]{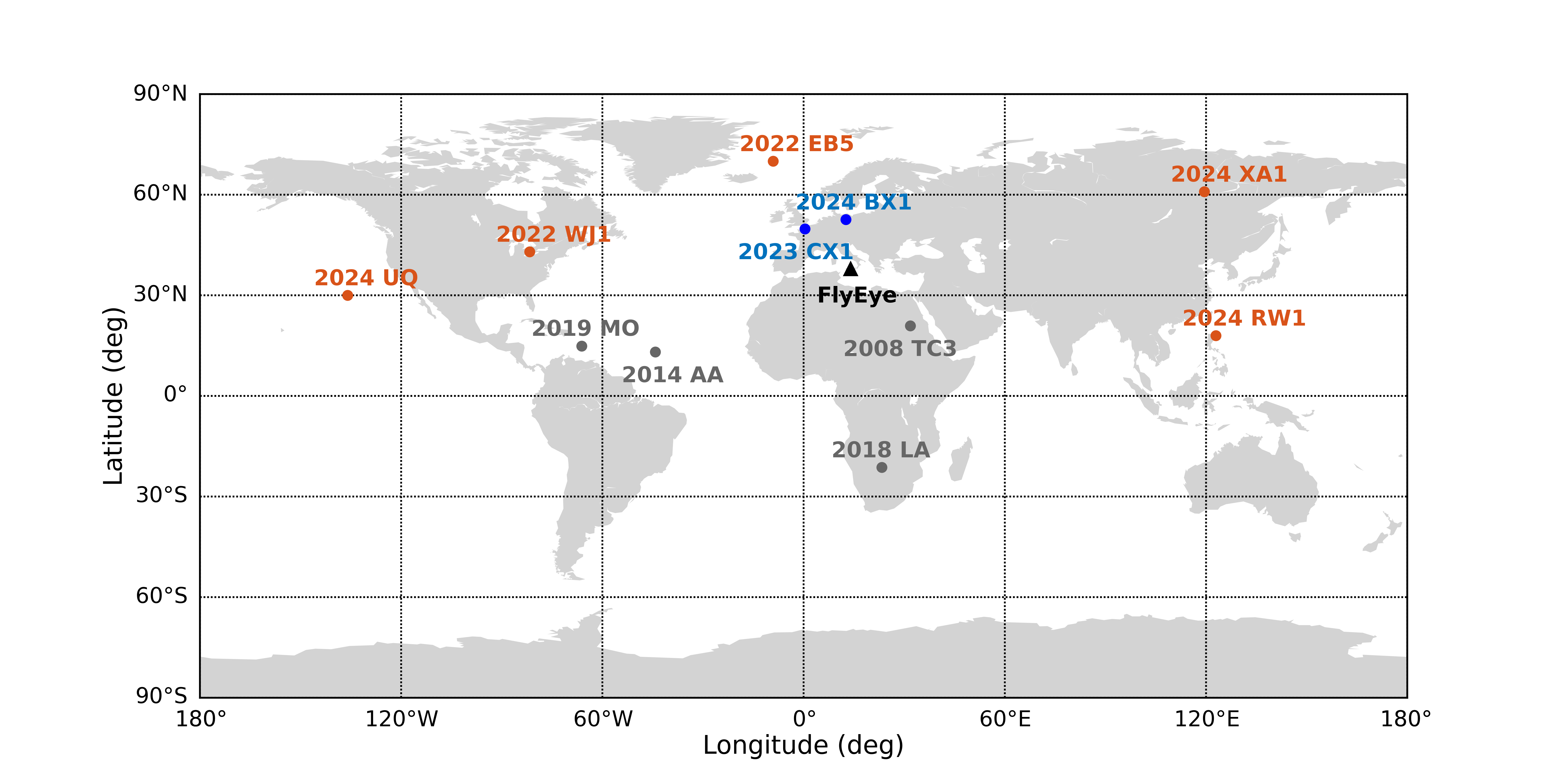}
    \caption{Impact locations of the imminent impactors detected by Meerkat are shown in blue and red. Red dots denote the cases for which meteorites were not found, while blue dots are the cases for which meteorites were recovered. For completeness, the dark gray dots indicate the imminent impactors which happened before the beginning of Meerkat's operational lifetime.}
    \label{fig:impact_location}
\end{figure}

Since operations commenced in 2021, Meerkat identified and sent alarms for several interesting cases, including imminent impactors and close approaches with Earth. Fig. \ref{fig:impact_location} shows the impact locations for all the imminent impactors detected before impact, identifying those that occurred during Meerkat's operational lifetime.  Here we summarise some of these cases, highlighting the role of Meerkat in triggering the confirmation and follow-up of these objects. An overview is given in Tables \ref{tab:impactors} and \ref{tab:close_approaches}, with a more detailed story for each object written in Sections \ref{subsec:past_impactors} and \ref{subsec:close_approaches}.

\begin{table}[!ht]
    \centering
\begin{tabular}{l c l l c l c l c}
\hline
Desig & Disc. stn & $t_{\text{rep}}$ &
$t_{\text{upd}}$ & $n$ &$t_{\text{alert}}$ & Imp (\%) & $t_{\text{imp}}$  & $\sim$Size (m)\\
\hline
2022~EB$_5$ & K88 & 19:38 &20:16 &14& 20:23 &100.0&21:23 & 1-3 \\
2022~WJ$_1$ & G96 & 05:31 & 05:31 &4& 05:36 &21.3  &08:27 & 0.5-1\\
2023~CX$_1$ & K88 & 20:49&21:29 &7& 21:33 &100.0 &02:59$^{+1}$ & 1-2\\
2024~BX$_1$ & K88 &22:15 &22:33 &7& 22:36 &100.0 & 00:32$^{+1}$& 1-2\\
2024~RW$_1$ & G96 &06:39 & 06:39&4& 06:48 &20.3 & 16:40& 1-2\\
2024~UQ     & T05 &19:41& 19:41&6& 19:44 &99.8 & 10:54 & 1-2\\
2024~XA$_1$ & V00 &07:05 &07:46&8& 07:50 &100.0&16:15 & 0.5-2\\
\hline
\label{tab:impactors}
\end{tabular}
\caption{Summary of impacting objects which triggered Meerkat alerts in the past. The designation (desig) of each object is given, along with the discovery station (Disc. stn), time of first being reported on the NEOCP ($t_{\text{rep}}$), time of first NEOCP update to trigger a Meerkat alert with impact probability above 10\% $(t_{\text{upd}})$, number of observations at $t_{\text{upd}}$ ($n$), the impact score (Imp), the time of impact ($t_{\text{imp}}$) and the approximate size. All times are given in UTC, with a superscript $({+1})$ indicating the following day. On many occasions, the first NEOCP report was also the first set of observations to trigger a Meerkat alert with an impact score above the impact alert threshold. }
\end{table}

\begin{table}[!ht]    
\centering

\begin{tabular}{l c l l c l c l c}
\hline
Desig & Disc. stn & $t_{\text{rep}}$ &
$t_{\text{upd}}$ & $n$ &$t_{\text{alert}}$ & $d_{\text{CA}}$ (km) & $t_{\text{CA}}$ & $\sim$Size (m)\\
\hline
2023~BU    & L51 & 00:46& 00:46& 3& 00:52& 3606& 00:27$^{+6}$&4-5\\  
2024~LH$_1$& 703 & 10:10& 10:30& 8& 10:34& 1700& 14:01       &1-3\\  
2024~XA    & 703 & 05:38& 06:06& 7& 06:10& 1322& 09:46       &1-2\\  
2024~UG$_9$& G96 & 07:43& 07:43& 3& 07:49& 2360& 12:42       &1-2\\  
\hline
\label{tab:close_approaches}
\end{tabular}
\caption{Summary of close approaching objects which triggered Meerkat alerts. The designation (desig) of each object is given, along with the discovery station (Disc. stn), time of first being reported on the NEOCP ($t_{\text{rep}}$), time of first NEOCP update to trigger a Meerkat alert with the weighted mean distance of close approach less than 7 Earth radii ($t_{\text{upd}}$), number of observations at $t_{\text{upd}}$ ($n$), the distance of close approach from the Earth's surface ($d_{\text{CA}}$), the time of close approach ($t_{\text{CA}}$) and the approximate size. All times are given in UTC, with a superscript $({+x})$ indicating the number of days following ($t_{\text{rep}}$). In some cases, the first NEOCP report was also the first set of observations to trigger a Meerkat alert meeting the close approach alert threshold.}
\end{table}

\subsection{Past impactors}\label{subsec:past_impactors}

\paragraph{2022 EB$_5$}
The first alert of a certain impact issued by Meerkat was for the object Sar2593 (designated as 2022~EB$_5$). The alert was triggered on 11 March 2022 at 20:23 UTC when observations from the Konkoly Observatory in Hungary were submitted to the NEOCP. At this time, the impact location was already constrained to about 100 km with an impact time between 21:21 and 21:25 UTC, so only one hour after the first alert. Thanks to follow-up observations obtained by the alert, the time and location of the impact could be better constrained. The object entered the atmosphere at approximately 21:23 UTC at the predicted area, located a few hundred kilometres north of Iceland. 
More details regarding this object can be found in \citet{geng-etal_2023}.

\paragraph{2022 WJ$_1$}
\label{subsubsec:2022WJ1}
The second alert of an imminent impactor was for object C8FF042 (designated as 2022~WJ$_1$), reported first by the Mount Lemmon Survey, on 19 November 2022 at 05:31 UTC. With only the first 4 observations reported by the NEOCP page, Meerkat determined an impact score of about 20\%. The impact corridor computed with this data indicated a most-likely hitting point in North America. Within minutes of the notifications, observers started obtaining follow-up observations of the new unconfirmed asteroid. In less than 30 minutes from the initial trigger, the impact became confirmed thanks to newly reported observations. The impact time and location could also be refined: the object was going to impact somewhere between Lake Erie and Lake Ontario, near the USA-Canada border, around 08:27 UTC. At exactly the predicted time, a fireball event happened at the expected location. The event likely dropped meteorites, however the current estimate indicate that they might have ended in Lake Ontario \citep{carbognani-etal_2025}.
More details about this object can be found in \citet{kareta-etal_2024}.

\paragraph{2023 CX$_1$}
This 1 m asteroid was discovered by K. Sárneczky from the Konkoly Observatory in Hungary. At 20:18 UTC on 12 February 2023 the object was imaged by Sárneczky, followed by a second position reported at 20:49 UTC. The object was given the temporary designation Sar2667 (designated as 2023~CX$_1$) on the NEOCP. About 40 minutes later, some follow-up observations reported by the Višnjan Observatory in Croatia confirmed the object, and at this point the imminent impactor monitoring systems found a 100\% impact score in the area of the English Channel between 02:00-04:00 UTC. The estimated asteroid size was around 1 m, and posed no risk of damage for the people in the area. During the following seven hours, astronomers all around the globe observed the object and pinpointed the impact corridor at 50 km altitude over the English Channel, with a trajectory coming from West to East. The object was observed up to around 10 minutes before impact, only 5 minutes before entering the Earth shadow and becoming unobservable. The last image was taken at 02:46 UTC, by ESA collaborators of the Rantiga Observatory in Italy. The fireball event happened at the predicted time (02:59 UTC) and location, with observations mostly from Southern UK and France, but also from Belgium, the Netherlands and Germany. Meteorite search campaigns were organized in the days immediately after the fall, and fragments were retrieved near the village of Saint-Pierre-le-Viger (France) \citep{bischoff-etal_2023}. %The object that was later designated as 2023~CX$_1$ by the MPC.

The ESA Planetary Defence Office contributed to the event both with the timely notifications from the Meerkat system, and network of optical telescopes for immediate follow-up. Astrometry from South Africa, quickly after the initial trigger, and later from the USA, extended the observational baseline to continents outside Europe, providing larger parallax. In the few minutes before impact, they triggered accurately-timed observations useful to reduce uncertainty of the impact time to less than 1 second. 

\paragraph{2024 BX$_1$}
This asteroid was also discovered by K. Sárneczky. The discovery image was taken on 20 January 2024 at 21:48 UTC, and the first set of 3 astrometric measurements was submitted to the MPC about 30 minutes later. The object received the temporary designation of Sar2736 (designated as 2024~BX$_1$) on the NEOCP. In the next 20 minutes Sárneczky submitted 4 additional observations, enabling Meerkat to start the impact monitoring routines. The very first email alert was sent out to Meerkat subscribers at 22:36 UTC, while the NEOCC team was also notified by text messages. With only 7 observations, the impact score computed by Meerkat was already 100\%, with an impact location between Sweden and Germany and an impact time between 23:54 UTC to 02:31 UTC on the 21 January 2024. A few minutes later, notifications were also sent by Scout and NEOScan. 

After these alerts, the observing community reacted quickly to perform follow-up imaging and submit measurements to the MPC. The second Meerkat alert was sent at 22:53 UTC using 10 observations: the impact was still certain, the impact corridor shrank significantly, providing an estimate of the fall area somewhere 50 km west of Berlin (Germany). The impact time was also constrained between 00:26 UTC and 00:36 UTC on 21 January 2024. With 20 observations available at 23:23 UTC, the impact time computed by Meerkat had an uncertainty of about 1 minute. Over the following couple of hours, more than a dozen observatories in mainland Europe obtained follow-up observations. The ESA NEOCC team also observed the asteroid from Tenerife until the latest possible instant at 00:25 UTC, when the asteroid entered the shadow of the Earth and disappeared from view. These were the only observations outside mainland Europe, which added a significant parallax and greatly improved the orbit determination. Efforts in getting the lightcurve were also pursued \citep{devogele-etal_2024}, enabled by the early warnings. The impact event happened at the predicted time and location of 00:32 UTC, near Nennhausen, Germany. The fireball was filmed by webcams in Berlin and Leipzig, as well as local people who knew about the event and shared their footage afterwards on social media. %This object finally received the designation of 2024~BX$_1$ by the MPC.
The fireball was also caught by meteor cameras from the European Fireball Network and the AllSky7 network, which permitted an accurate strewn field to be computed for possible meteorites dropped during the atmospheric entry \citep{spurny-etal_2024}. Search campaigns were promptly organized, and meteorites fragments were recovered in the days following the impact near the village of Ribbeck. 

\paragraph{2024 RW$_1$}
\label{subsubsec:2024RW1}
This object was discovered by J. Fazekas from the CSS on 4 September 2024, and received the NEOCP designation CAQTDL2 (designated as 2024~RW$_1$). With only 4 observations, the object had an impact score of 20.3\%, and the Meerkat alarm was sent out to the mailing list at 06:48 UTC. The NASA Scout system also sent an alert a few minutes before Meerkat, at 06:42 UTC.
Follow-up observations were immediately performed by CSS and from the Magdalena Ridge Observatory in Socorro. With 8 observations the object already reached 100\% probability of impact, and had an estimated size of about 1 meter. The impact corridor stretched over a few hundred kilometres, with part of it over the island of Luzon (Philippines). The predicted impact time was around 16:40 UTC, corresponding to near midnight local time. New observations helped improve the determination of the impact location, which shifted towards the western Pacific Ocean. Photometric observations yielded a double-peaked rotation period of $\sim$1900 s and  the closest spectral type match being to B-type asteroids \citep{ingebretsen-etal_2025}. %The object was designated as 2024~RW$_1$ by the MPC.

\paragraph{2024 UQ}
Object A11dc6D (designated as 2024~UQ) was discovered by one of the ATLAS telescopes in Hawaii, roughly 2 hours prior to impact. Unfortunately, submission to the NEOCP only happened after the predicted impact time. Still, with only 6 observations, Meerkat computed a certain impact and sent out an alert to the mailing list. Since in this event there was no option for following-up the object, the astronomical community searched for precoveries to confirm the object as an asteroid. 
The object was found in images from CSS obtained a few minutes before the ATLAS discovery observations, and helped in improving the computation of the atmospheric entry location. A flash was also recorded by the Geostationary Operational Environmental Satellite, providing a first confirmation of the impact.
Later on, the NASA CNEOS reported a fireball at coordinates 30.0 N 136.0 W at 10:54:48 UTC on the 22 October 2024, compatible with the impact trajectory from the observational data. The combined evidence was sufficient to confirm an impact over the Pacific Ocean.

\paragraph{2024 XA$_1$}
Object C0WEPC5 (designated as 2024~XA$_1$) was discovered by V. F. Carvajal from the Bok Telescope at Kitt Peak on 3 December 2024. About one hour after the first observation, the object was imaged by the Steward Observatory at Mount Lemmon, and at 07:46 UTC a total of 8 observations were posted on the NEOCP. The Meerkat impact alert was sent out at 07:50 UTC, indicating a 100\% impact score between roughly 15:00 - 17:00 UTC. The estimated size from these first observations was only 0.6 m. The initial impact corridor was located over the Sahka Republic (Russia). Before impact, a total of eight updates were posted on the NEOCP, and an impact alarm continued to be issued. At the time of the last update, the impact time predicted by Meerkat was 16:15 UTC. The fireball was caught by eyewitnesses and a live webcam located in the city of Lensk (Russia), right at the predicted time. 
Most likely, meteorite fragments were dropped during the atmospheric entry, and details of the predicted strewn field were provided \citep{gianotto-etal_2025}. However, the predicted location is covered by taiga vegetation and, as of the time of writing this paper, no meteorite search campaign has been successful.

\subsection{Close approaches}\label{subsec:close_approaches}

\paragraph{2023 BU}
On 22 January 2023, G. Borisov\footnote{An amateur astronomer who also discovered the first interstellar comet 2I/Borisov.} reported the observations of an object which was assigned the designation of gb00553. Meerkat sent an impact alarm with a probability of 53\%, and the computations were based on the first 6 reported astrometric measurements covering an arc of 1.2 h. The size was estimated to be about 5 m in diameter, and the estimated impact time spanned an interval of almost two days. The impact corridor was almost entirely contained at latitudes between 0 and 40 deg wrapping around the Earth, thus covering all the longitude values. When 3 other observations were reported on the NEOCP, the arc length increased to 2.14 h, the orbit was better constrained and the impact score dropped to 21\%. With 6 more  measurements and an arc length of 4.26 h the impact chances were completely ruled out by Meerkat. On the same day the MPC announced the discovery of this asteroid, it received the designation of 2023~BU. 
The observational data were also ingested by the ESA NEOCC Aegis orbit determination and impact monitoring system \citep{fenucci-etal_2024}, which confirmed a zero impact score during the upcoming close approach. The flyby of 2023~BU with Earth occurred on 27 January 2023 at 00:27:10 UTC, passing at 3606 km from Earth's surface at a speed of 9.26 km s$^{-1}$. This was the fourth closest approach ever at the time of the discovery epoch.

\paragraph{2024 LH$_1$}
In the afternoon of 6 June 2024 Meerkat sent an alarm of the NEOCP object C43AUL1 (designated as 2024~LH$_1$), which had 8 observations and no impact score. However, it notified a close approach at about 1.28 Earth radii, about 1700 km above the Earth's surface. Meerkat also gave a 100\% score for this object to be of NEA origin, with a size of about 2 m. Discovery observations were submitted by the CSS to the NEOCP, and the object was promptly followed up by NEOCC astronomers after the Meerkat notification from stations T03, T04, and F65 in Haleakala, Hawaii. Follow-up observations confirmed the object was real and confirmed the distance of the close approach. %The MPC later designated this asteroid as 2024~LH$_1$, and it is currently the third closest approach ever recorded.
This was the second closest approach ever recorded at the time of the discovery epoch.

\paragraph{2024 XA}
Object C43XT81 (later designated as 2024~XA) was first observed by CSS on 1 December 2024, and with a tracklet of 7 observations Meerkat issued an impact alert at 06:10 UTC, with a probability of 22\%. The first estimate for the size was about 1 m, similar to the other impacted cases. Immediate follow-up observations from the Steward Observatory in Mount Lemmon confirmed the object, and at the same time the impact was ruled out 30 minutes after the alert. 
It safely flew past Earth at a distance of only 1322 km from the Earth surface, the second closest approach ever recorded at the time of the discovery epoch.

\paragraph{2024 UG$_9$}
Object CC7E1A2 (later designated as 2024~UG$_9$) was discovered by CSS, and a set of 3 observations was reported and posted to the NEOCP in the early morning UTC on 30 October 2024. With this tracklet Meerkat computed an impact score of 15\%, and soon after JPL Scout confirmed this result. The first alarm was sent at 07:49 UTC. A follow-up campaign was promptly triggered, and about an hour after the first alert, the probability computed by Meerkat dropped to 5.9\% with 12 observations. Astrometric measurements performed by NEOCC from the V39 observatory were ingested by Meerkat at 08:14 UTC, and impact solutions were completely removed. This object had a flyby at 2360 km from Earth surface, the third smallest close approach at the time of the discovery epoch.

\section{Spurious alerts}
Objects can frequently appear in the NEOCP with observational arcs of less than 0.1 h. Such extremely short arcs can be a problem even for systematic ranging due to the lack of curvature. At such an arc length, virtually all $(\rho_i, \dot{\rho}_j)$ in the grid are fitted equally well. This is made apparent by the  lack of variation in the WRMSE, as quantified by the standard deviation of WRMSE values, $\sigma_{\text{WRMSE}}$. Consequently, the error function $f_{\text{err}}$ becomes uniform and the posterior probability tends towards the prior distribution $f_\text{prior}$. This can inflate the impact score to non-negligible values, occasionally over the 10\% threshold required to trigger alerts from the Meerkat system. Table \ref{tab:false-positives} lists some examples of false positive alerts issued by Meerkat over the past year. It can be seen that compared to a typical value of $\sigma_{\text{WRMSE}}$ of $\sim 10^{3}$ for low observations, the very small standard deviation confirms the degeneracy of the grid fitting from systematic ranging. An exception to this is SVO005, which showed a high $\sigma_{\text{WRMSE}}$ value of 2150.74. This is blamed on poor observations, as ultimately the object did not exist. Still, the fact that we can see false positives with high impact scores above 10\% means these objects need to be treated carefully in the analysis. 

\begin{table}[!ht]
    \centering
\begin{tabular}{l c c c c c l}
\hline
Obj & $n$ & Imp (\%) & Arc length (h) & $\sigma_{\text{WRMSE}}$ & NEOCP removal \\
\hline
5C20001  & 4 & 0.9 & 0.04 &  0.11 & C then D \\
A11fW5A  & 5 & 0.3 & 0.04 &   0.04 & D \\
A11jqBd  & 5 & 1.9 & 0.04 &  0.22 & C then D \\
A11pTYb  & 4 & 5.0 & 0.04 &  0.07 & C then D \\
CARHLK2  & 3 & 1.5 & 0.05 &  0.08 & D \\
GHN0878  & 3 & 66.7 & 0.10 & 2.36 & L \\
gb00763  & 4 & 1.7 & 0.04 & 0.06  & D \\
% ML13578  & O18 & 8 & 100.0 & 1.10 & 3.06 & DNE \\ repeated obs
SVO005   & 3 & 100.0 & 0.17 & 2150.74 & DNE \\
ZTF103e  & 6 & 21.2 & 0.09 & 0.11 & D \\
ZTs0374  & 4 & 3.5 & 0.02 & 0.03 & D \\
ZTs0379  & 4 & 4.3 & 0.02 & 0.01 & L \\
ZTs0386  & 4 & 4.9 & 0.02 & 0.01 & L \\
ZTs0403  & 4 & 2.2 & 0.02 & 0.02  & D \\
ZTs0415  & 4 & 4.2 & 0.02 & 0.03 & D \\
\hline
\label{tab:false-positives}
\end{tabular}
\caption{Examples of false positive alerts issued between October 2024 and August 2025. The number of observations ($n$), impact score (Imp), arc length and standard deviation of the grid of WRMSE values ($\sigma_{\text{WRMSE}}$) are given for each object. In addition, the reason for the object being removed from the NEOCP is indicated, with flags C (combined), D (designated), L (lost) and DNE (does not exist). All of the objects presented have an arc type of 1, as defined in \cite{milani-etal_2007}.}
\end{table}

\begin{figure}[!ht]
    \centering
    \includegraphics[width=\textwidth]{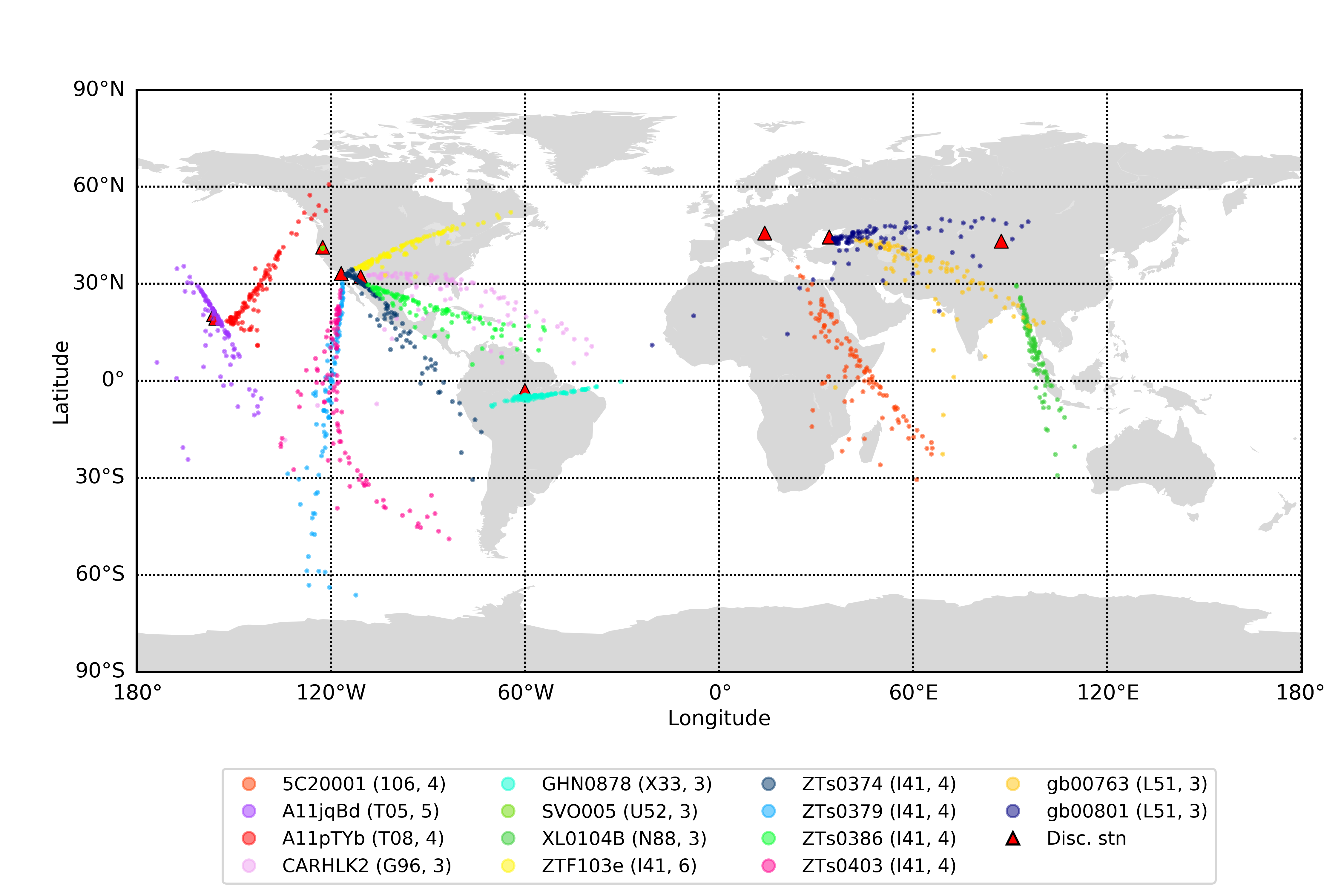}
    \caption{Examples of the impact corridors derived from Monte Carlo sampling for some of the false positive alerts in Table \ref{tab:false-positives}. A total of 100 heliocentric impacting samples are drawn for each object. The discovery station and number of observations are indicated, with each of the discovery stations plotted on the map as a red triangle.}
    \label{fig:imp-line-towards_stn}
\end{figure}

Many of these false positive alerts form an unusual shape in the Meerkat impact plot; impact locations appear to be concentrated in a line pointing towards the discovery station. Examples of this behaviour are shown in Fig. \ref{fig:imp-line-towards_stn}. 

\begin{figure}[!ht!]
    \centering
    \includegraphics[width=\textwidth]{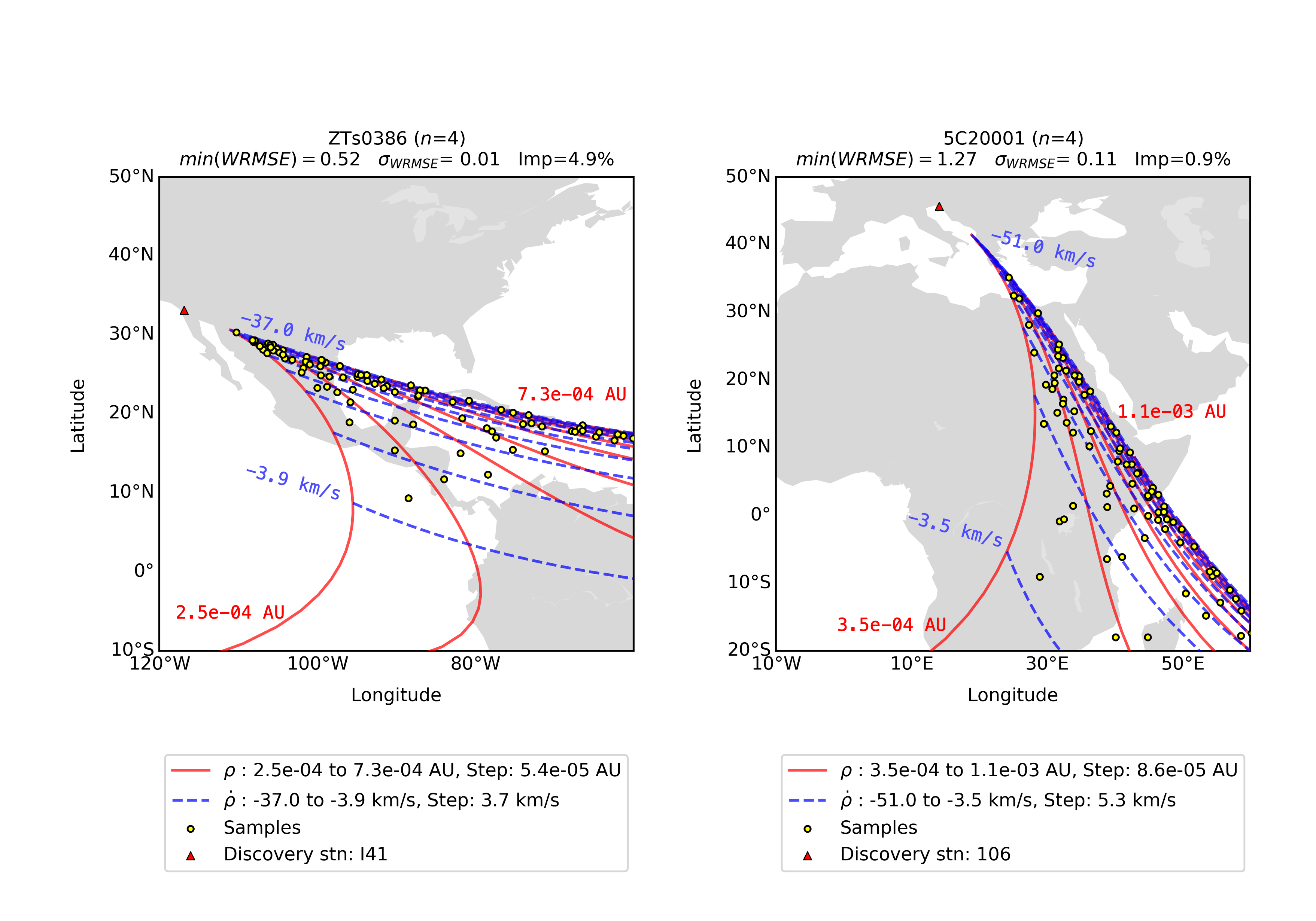}
    \caption{The mapping of the systematic ranging mesh in topocentric range, $\rho$ (red), and range rate , $\dot{\rho}$ (blue) space onto latitude - longitude space. Monte Carlo samples corresponding to 90 heliocentric impacting solutions are plotted. The minimum WRMSE, standard deviation of the WRMSE ($\sigma_{\text{WRMSE}}$) and impact probability are also indicated.}
    \label{fig:grid_mapping_impact}
\end{figure}

This unusual impact corridor is a direct result of the flattened posterior probability distribution. The Monte Carlo samples are drawn from an approximately uniform distribution, so they are scattered more or less evenly across the impacting region of the grid. Consequently, the sample impact locations trace the mapping of the systematic ranging grid in $(\rho, \dot{\rho})$-space onto the impact plot in (lat, lon)-space on Earth surface. Two examples of this are given in Fig. \ref{fig:grid_mapping_impact} for objects ZTs0386 and 5C20001. 

Where the density of the mapping manifold is highest is also where the density of sample impact locations is highest. The density increases towards the negative range rate limit, due to the spatial dispersion of samples with different fitted velocities. This effect can be worsened by slower range rate samples having more time to be perturbed by Earth’s gravity, and typically being fitted with less eccentric orbits. The grid mapping converges at a point near the discovery station. This is an artificial limit, corresponding to an edge of the systematic ranging grid between the impacting region and the unbound region, with a low range and large negative range rate. The vicinity to the discovery station arises from the fact that the topocentric range and range rate are defined from the first observation station corresponding to epoch $t_1$. 

It should be noted that a suspected false positive alert should be identified from the observation arc length, curvature and arc type, and that this impact plot shape is only a consequence of these parameters. Such spurious alerts need to be treated carefully by analysts and observers. To aid in their identification, the standard deviation of the grid of WRMSE values is included in the Meerkat email, as a metric to quantify the degenerate fitting solutions. Additionally, a warning is issued at the top of the email for all objects with arc lengths less than 0.1 h.

\section{Summary}
This paper introduces the Meerkat Asteroid Guard, an independent, fully automated system that delivers alerts for imminent impactors and other scientifically interesting objects. 
The version 2.0 of the software underwent a significant redesign, introducing  \texttt{python3.11} and the GODOT (v1.9.0) flight dynamics library, as well as transitioning to modern development paradigms. 

Meerkat scans the NEOCP list 24 hours a day, 7 days a week to look for new objects and observation updates. Orbit determination is performed via systematic ranging over a large grid of topocentric range and range rates. 
The posterior probability for each grid point is computed from the WRMSE values and a uniform prior distribution. Each grid point has its absolute magnitude fitted and trajectory propagated forward 30 days to monitor potential impacts or close approaches. In addition, Monte Carlo samples are drawn and propagated to assess potential impact locations and derive a recommended telescopic pointing for follow-up observations. 

The grid of fitted orbital elements with their associated posterior probability provides a statistical description of the object's likely orbital and physical characteristics. These are presented as a set of posterior-weighted scores, including the impact probability, location and time, close approach distance, absolute magnitude, estimated size, orbital class and more. A series of informative plots are produced to display an overview of the object scores, the output of the systematic ranging grid, the detection probability as a function of time and field of view, the drawn Monte Carlo heliocentric and/or geocentric orbital elements compared to those of natural and artificial object classes and finally the impact corridor on Earth. 

Objects with a given impact score, close approach distance, ISO score or IEO score will trigger alerts issued as emails to Meerkat subscribers. These alerts have proved to be extremely beneficial to the astronomical community, aiding in the classification, monitoring and follow-up of thousands of potentially hazardous or scientifically interesting objects in the NEOCP. Since its launch in 2021, Meerkat has successfully provided warnings for all six past imminent impactors detected before impact. Notably, for all but one of these objects, Meerkat was the first of the existing asteroid warning systems to raise the alarm. 

With the publishing of this paper comes the release of Meerkat version 2.0. The upgrade has transformed the system architecture, computation speed and error rate, as well as providing several new features to aid observers and analysts. 
Additional features and improvements are foreseen for the near future. 
Frequent false positive alerts continue to be produced from poor observations or observations lacking arc length and curvature. An outlier detection algorithm is planned to be introduced in the future, to help identify erroneous observations. Implementing this with the method of systematic ranging may prove a statistical challenge and yield a computational cost, due to the nature of performing $N^2$ individual iterative least squares orbit determination routines, so would require some careful planning to develop. 

Additionally, the observation weighting scheme is planned to be upgraded to include the observer-defined uncertainty. From observations in the 80-column format, the uncertainty in right ascension and declination are derived from the weighting scheme of \cite{veres-etal_2017}. Using the MPC PostgreSQL database or an Astrometry Data Exchange Standard (ADES) data format, observer-defined values for $\sigma_{\alpha}, \sigma_{\delta}$, their correlation and $\sigma_{t}$ can be incorporated into the weighting scheme directly. %The implementation is planned to follow the approach of the ESA Aegis software \citep[][]{fenucci-etal_2024}, in which each weight is determined as the magnitude of a vector formed by two components: the observer-defined weighting and the scheme-defined weighting.
Finally, the use of graphics processing units (GPUs) for increasing the computational speed will be investigated, further exploiting the parallel nature of computations.
Meerkat has served as a reliable and informative imminent impactor warning service for the past five years. While the past imminent impactors have all been relatively small, $\sim$1 m in diameter, there remains the potential for an impactor of more significant size, $\sim$10 m or greater, to be discovered with little warning time. In such a scenario, the rapid threat assessment and suggested follow-up targeting provided by Meerkat would be greatly beneficial, helping to ensure public safety. 
As new observing facilities and surveys like the ESA Flyeye telescope and the Vera Rubin Observatory come into operations, the number of objects submitted to the NEOCP each day is expected to increase \citep{ivezic-etal_2006, grav-etal_2016}. Meerkat continues to be developed and improved over time, ready to respond to the next imminent impactors of the future.

\bibliographystyle{apalike85}
\bibliography{holybib}

\end{document}